\documentclass[aps,pra,amsmath,amssyb,floatfix,
twocolumn,superscriptaddress,10pt]{revtex4-1}
\usepackage{graphicx}
\usepackage{subfigure}
\usepackage{hyperref}
\usepackage{braket}
\usepackage{bm}

\def\124{YBa$_2$Cu$_4$O$_8$ }

\def\C60{A$_x$C$_{60}$ }

\def\TMTSF{(TMTSF)$_2$X }

\def\ie{{\it i.e.}}
\def\eg{{\it e.g.}}
\def\bnabla{\bm{\nabla}}

\def\br{{\bf r}}
\def\bn{{\bf n}}
\def\bv{{\bf v}}

\def\bk{{\bf k}}

\def\bJ{{\bf J}}
\def\bE{{\bf E}}

\def\H{\mathcal{H}}
\def\J{{\mbox{\small$\cal J$}}}
\def\sJ{\scriptstyle{\cal J}}
\def\Tr{{\rm Tr}}
\def\Lchi{{\mbox{\large$\chi$}}}
\def\prl{{Phys. Rev. Lett. }}
\def\prb{{Phys. Rev. B }}

\def\be{\begin{equation}}
\def\ee{\end{equation}}
\def\ba{\begin{eqnarray}}
\def\ea{\end{eqnarray}}

\begin{document}

\title{Transverse Thermoelectric Response as a Probe for Existence of Quasiparticles}

\author{Yoni~Schattner}
\affiliation{Racah Institute of Physics, The Hebrew University, Jerusalem 91904, Israel}
\affiliation{Department of Condensed Matter Physics, Weizmann Institute of Science,
Rehovot 76100, Israel}
\author{Vadim~Oganesyan}
\affiliation{Department of Engineering Science and Physics,
College of Staten Island, CUNY, Staten Island, NY 10314, USA}
\affiliation{Initiative for the Theoretical Sciences, The Graduate Center, CUNY,
New York, NY 10016, USA}
\author{Dror~Orgad}
\affiliation{Racah Institute of Physics, The Hebrew University,
Jerusalem 91904, Israel}

\date{\today}

\begin{abstract}

The electrical Hall conductivities of any anisotropic interacting system with reflection symmetry
obey $\sigma_{xy}=-\sigma_{yx}$. In contrast, we show that the analogous relation between the
transverse thermoelectric Peltier coefficients, $\alpha_{xy}=-\alpha_{yx}$, does not generally
hold in the same system. This fact may be traced to interaction contributions to the heat current
operator and the mixed nature of the thermoelectric response functions. Remarkably, however,
it appears that emergence of quasiparticles at low temperatures forces $\alpha_{xy}=-\alpha_{yx}$.
This suggests that quasiparticle-free groundstates (so-called non-Fermi liquids) may be detected by
examining the relationship between $\alpha_{xy}$ and $\alpha_{yx}$ in the presence of reflection
symmetry and microscopic anisotropy. These conclusions are based on the following results: (i) The relation
between the Peltier coefficients is exact for elastically scattered noninteracting particles; (ii) It holds
approximately within Boltzmann theory for interacting particles when elastic scattering dominates over
inelastic processes. In a disordered Fermi liquid the latter lead to deviations that vanish as $T^3$.
(iii) We calculate the thermoelectric response in a model of weakly-coupled spin-gapped Luttinger
liquids and obtain strong breakdown of antisymmetry between the off-diagonal components of $\hat{\alpha}$.
We also find that the Nernst signal in this model is enhanced by interactions and can change sign as
function of magnetic field and temperature.

\end{abstract}

\pacs{72.15.Jf,73.50.Lw,65.90.+i,71.10.Pm,74.40.-n}

\maketitle

\section{Introduction}
\label{sec:inrto}

Typically, an electronic system sustains average charge and heat current densities,
${\bm J}^e$, ${\bm J}^h$, when subjected to a uniform temperature gradient, ${\bm \nabla} T$,
and constant electric field, $\bE$. Its linear thermoelectric response is described by
\be
\label{eq:linres}
\left( \begin{array}{c}
{\bm J}^e\\
{\bm J}^h \\ \end{array}\right) =
\left(\begin{array}{cc}
\hat\sigma & \hat\alpha\\
\hat{\tilde\alpha} & \hat\kappa\\ \end{array}\right) \left(\begin{array}{c}
\bE \\
-{\bm \nabla T}\\ \end{array}\right),
\ee
where $\hat\sigma$ is the conductivity tensor, $\hat\alpha$ and $\hat{\tilde \alpha}$ are the Peltier tensors,
and $\hat\kappa$ is the thermal conductivity tensor.
In noninteracting systems, the electrical and heat-current operators are simply related to each other,
giving rise to relations between $\hat\sigma$, $\hat\kappa$ and $\hat\alpha$. These relations
continue to hold in Fermi liquids, up to asymptotically vanishing corrections. An example is the
Wiedemann-Franz law, $\hat\kappa= (\pi^2/3e^2)T\hat\sigma$, (we use throughout $\hbar=k_B=1$. $-e<0$
is the electron charge), whose breakdown has been interpreted as a signature of physics beyond the Fermi liquid
framework \cite{Aleiner,Karen-kinetic,Orignac-EPL,Efrat-W-F}. Another is the exact relation for
noninteracting electrons \cite{Streda77,Jonson} between $\hat\alpha$ at a given temperature $T$ and chemical
potential $\mu$, and $\hat\sigma$ of the same system at zero temperature and shifted chemical potential
\be
\label{eq:Mott}
\hat\alpha(T,\mu) = \frac{1}{e T} \int_{-\infty} ^{\infty} d\epsilon\, \epsilon
\frac{\partial n_F(\epsilon)}{\partial \epsilon}\hat\sigma(T=0,\mu+\epsilon),
\ee
where $n_F(\epsilon)$ is the Fermi function.
This formula hence implies that in the absence of interactions, $\hat\alpha$
shares the same symmetry properties as $\hat\sigma$. A similar conclusion is reached by solving the
Boltzmann equation within an energy-dependent relaxation-time approximation \cite{Ziman,Abrikosov}.

Owing to the pioneering works of Onsager \cite{Onsager} and subsequently of Kubo \cite{Kubo} it is well known
that various linear-response transport coefficients are related via the time reversal symmetry of microscopic
dynamics. Consequently, one finds on general grounds that in the presence of a magnetic field $B$,
$\sigma_{ij}(B)=\sigma_{ji}(-B)$ and $\tilde\alpha_{ij}(B)=T\alpha_{ji}(-B)$, where $i,j=x,y,z$. In turn,
it is straightforward to show that even for an anisotropic system, as long as it is invariant under reflections,
say with respect to the $y$ axis, $\sigma_{xy}(B)=-\sigma_{yx}(B)$. The above discussion implies that under similar
conditions one also finds $\alpha_{xy}(B)=-\alpha_{yx}(B)$, provided that the system is noninteracting or considered
within approximated Boltzmann transport theory. A natural question then arises: Is the relation
$\alpha_{xy}(B)=-\alpha_{yx}(B)$ valid beyond the limits of these two conditions? Beside its intrinsic
theoretical appeal, this issue is also important for identifying non-Fermi liquid behavior in the thermoelectric
properties of correlated electronic systems.

One such property is the Nernst signal, defined by the off-diagonal elements $S_{xy}$ and $-S_{yx}$
of the thermopower tensor $\hat S=\hat\sigma^{-1}\hat\alpha$. The latter relates the measured electric field to
an applied temperature gradient, $\bE=\hat S{\bm \nabla} T$, in the presence of a magnetic field
$B_z$ and in the absence of an electrical current. The dependence of $\hat S$ on both the resistivity tensor
$\hat\rho=\hat\sigma^{-1}$ and $\hat\alpha$ means that generally $S_{xy}= -S_{yx}$ only for isotropic systems.
Therefore, the symmetry properties of $\hat S$ do not carry direct information about interaction effects without
independent knowledge of $\hat{\sigma}$.
However, such information may be gleaned from discrepancies between the measured Nernst signal and the predictions
of Boltzmann transport theory. While this theory accounts for the observed data in a number of
materials \cite{Behnia-Review} it underestimates the effect by orders of magnitude in several quasi-one-dimensional
conductors \cite{Ong-Chaikin-1dNernst,1d-Nam,purple}.

The Nernst effect is also a sensitive probe of superconducting fluctuations, which contribute positively to the
signal \cite{Dorsey,Ussishkin1,Michaeli1,Podolsky,Raghu,Galitski,Levchenko}, in contrast to quasiparticles of
various ordered normal states whose contribution is often of a negative sign \cite{Ido-Vadim,Vojta}.
A positive Nernst effect has been measured over a wide range above the critical temperature, $T_c$,
in a series of superconductors including the
cuprates \cite{Ongxu,OngPRB01,Onglongprb,Taillefer-Nature-stripes,Taillefer-broken-symm-Nature,Taillefer-anis-PRB,Taillefer-Nature-Phys12},
as well as amorphous films of Nb$_{0.15}$Si$_{0.85}$ and InO$_x$ \cite{Nature-Phys-fluct,Pourret09}.
While the fluctuation contribution in the cuprates emerges from a high-temperature negative quasiparticle signal,
the latter dominates the Nernst effect down to, and even below, $T_c$ in other compounds such
as the pnictides \cite{Wang-Pnic,Matusiak-Pnic,Kondart-Pnic}. It is therefore interesting to investigate the
interplay between these opposing contributions in systems which exhibit concomitant strong fluctuations
towards competing orders including superconductivity.

Motivated by the aforementioned issues we study in Sec. \ref{sec:alpha} the symmetry properties
of $\hat\alpha$ within a generic model of interacting electrons.
We begin by considering the thermoelectric linear response using the
Kubo formula. We show that the close relation which exists between the electrical and heat current
operators in the noninteracting limit naturally leads, in the presence of reflection symmetry,
to $\alpha_{xy}(B)=-\alpha_{yx}(B)$. However, contrary to the corresponding relation for the Hall
conductivities the property $\alpha_{xy}(B)=-\alpha_{yx}(B)$ is not protected by reflection and
time-reversal symmetries, and we demonstrate its explicit violation in the exactly solvable problem
of two harmonically interacting electrons in a magnetic field. Having established this point of
principle we move on to consider the issue using Boltzmann transport theory for the interacting system.
We show that $\alpha_{xy}(B)=-\alpha_{yx}(B)$ is obtained within the relaxation-time approximation
of this theory, or more generally whenever inelastic processes can be neglected. Since this is the case
in a disordered Fermi liquid at low temperatures we conclude that violation of the above
relation under the specified conditions is a telltale sign of interactions beyond the Fermi liquid framework.

In Sec. \ref{sec:Nernst} we consider a non-Fermi liquid model of weakly coupled Luttinger chains in the presence
of a spin gap. We show that the antisymmetry of the off-diagonal elements of $\hat\alpha$
is indeed violated. Furthermore, we calculate the Nernst signal and show that interactions can lead
to its substantial enhancement in such low dimensional systems. This may bare
relevance to understanding the large signal observed experimentally in the quasi-one-dimensional
materials. Finally, we also find that the sign of the effect in the spin gapped system changes
from negative to positive as the temperature is lowered and the magnetic field increased. We
interpret this behavior as being due to the stronger superconducting fluctuations induced
by the spin gap. Various technical aspects of our study are relegated to the appendices.

\section{The symmetry properties of {\Large $\hat\alpha$}}

\label{sec:alpha}

\subsection{{\large $\hat\alpha$} within Kubo theory}
\label{subsec:Kubo}

We consider interacting spinless fermions in a two-dimensional system of area $A$,
which includes mass anisotropy and coupling to static electromagnetic potentials. The system is
described by the Hamiltonian $H=\int d^2r \H(\br)$, with
\ba
\label{eq:2dhdensity}
\nonumber
\H(\br)&=&\frac{1}{2m_\mu}\left[D_\mu\psi(\br)\right]^\dagger
\left[D_\mu\psi(\br)\right]-e\phi(\br)\rho(\br)\\
&&+\frac{1}{2}\int d^2r' U(\br-\br')\psi^\dagger(\br)\rho(\br')\psi(\br),
\ea
where $D_\mu=\partial_\mu+i(e/c)A_\mu(\br)$, summation over repeated Greek indices, which take the values $x,y$,
is implied, and the interaction is assumed to obey $U(\br-\br')=U(\br'-\br)$.

A route for calculating the thermoelectric coefficients was laid out by Luttinger \cite{Luttinger},
who argued that in the long-wavelength, low-frequency limit the linear response to a temperature
variation $\delta T({\bf r},t)$ is the same as the response to a fictitious gravitational field
$g({\bf r},t)=\delta T({\bf r},t)/T$. An extension of Luttinger's results to the case with
a magnetic field was given by Oji and Streda \cite{Oji}. The gravitational field enters the
calculation in two ways: First, it couples to the unperturbed density ${\cal K}={\cal H}-\mu\rho$
of $K=H-\mu N$, such that the latter reads $K_T=K+\int d{\bf r} g({\bf r},t){\cal K}({\bf r})$.
Secondly, the unperturbed current density operators $\bJ^e$, $\bJ^h$ are themselves
modified, with $\bJ^e$ becoming $\bJ^e+\delta\bJ^e=\bJ^e+g\bJ^e$, see Appendix \ref{app:currents}.
Consequently,
\ba
\label{eq:Kubo1}
\nonumber
\alpha_{ij}&=&\frac{1}{-\partial_{j}g}\frac{1}{AT}\left[
\left\langle\int d^2r {\rm J}^e_i ({\bf r})\right\rangle_{\!\! K_T} +
\left\langle \int d^2r \delta{\rm J}^e_i ({\bf r})\right\rangle_{\!\! K}
\right]\\
&&\equiv  \alpha_{ij}^{(1)} +\alpha_{ij}^{(2)}.
\ea
Henceforth, Latin indices, which take the values $x,y$, are not summed over, and
$\langle J\rangle_K ={\rm Tr}(e^{-\beta K} J)/Z_K$, where $\beta=1/T$, $Z_K={\rm Tr}(e^{-\beta K})$.

The contribution $\alpha_{ij}^{(2)}$ is analogous to the diamagnetic term in the electrical conductivity.
For a spatially constant temperature gradient one finds, (see Appendix \ref{app:Kubo})
\be
\label{eq:alpha2}
\alpha_{ij}^{(2)}=-\frac{1}{AT} \left\langle\int d^2r {\rm J}^e_i ({\bf r}) r_j \right\rangle_{\!\! K}
= \frac{c}{AT}\epsilon^{ijz}M_z,
\ee
where $M_z$ is the $z$ component of the orbital magnetization.
The importance of this contribution and its origin in the redistribution of the equilibrium magnetization
currents which flow in the system, has been extensively discussed by Cooper, Halperin and Ruzin \cite{Halperin97}.
Here, we note that its appearance is a direct consequence of the Kubo formalism.

Whereas $\alpha_{ij}^{(2)}$ is clearly antisymmetric in $i$ and $j$, the other contribution
(see Appendix \ref{app:Kubo})
\be
\label{eq:alpha1}
\alpha_{i j}^{(1)}=\lim_{\omega\rightarrow 0}\frac{A}{T} \frac{i}{ \omega+ i \delta}
\left[\Lchi_{J^e_i,J^h_j}(\omega+ i\delta)
-\Lchi_{J^e_i,J^h_j}(i\epsilon)\right],
\ee
expressed in terms of the retarded correlation function $\Lchi _{J^e_i,J^h_j}$ of the
averaged electrical and heat current densities, is generally not. The transformation properties
of the correlation functions are discussed in Appendix \ref{app:timerev}. Under spatial reflection,
when such a transformation is a symmetry of the system, they imply that the diagonal elements
of $\hat\alpha^{(1)}$ are even functions of the magnetic field ${\bf B}=B\hat {\bf z}$, while the
off-diagonal elements are odd. Since also $M_z(B)=-M_z(-B)$ one finds that
\be
\label{eq:refsymmalpha}
\alpha_{ij}(B) =
\left\{\begin{array}{cc}
\alpha_{ij}(-B) & i=j\\
-\alpha_{ij}(-B) & i\neq j\\ \end{array}\right. ,
\ee
with similar relations for ${\hat{\tilde\alpha}},\hat\sigma$ and $\hat\kappa$.

Concomitantly, the transformation of $\hat\alpha^{(1)}$ under time reversal and the
expressions for $\hat{\tilde{\alpha}}^{(1)}$ and $\hat{\tilde{\alpha}}^{(2)}$,
Eqs (\ref{eq:alphatilde1},\ref{eq:alphatilde2}), lead to the conclusion
\be
\label{eq:timerevsymmalpha}
T\alpha_{ij}(B) = \tilde{\alpha}_{ji}(-B).
\ee
Hence, combining property (\ref{eq:refsymmalpha}), when applied to $\hat{\tilde{\alpha}}$,
with Eq. (\ref{eq:timerevsymmalpha}) yields the relation $T\alpha_{ij}(B)=-\tilde\alpha_{ji}(B)$
between the off diagonal elements of the Peltier tensors. However, symmetry considerations
do not imply a similar relation between the elements of $\hat\alpha$, per se. This stands in
contrast to $\hat\sigma$ (and $\hat\kappa$), whose elements are related by time reversal
symmetry via $\sigma_{ij}(B)=\sigma_{ji}(-B)$, thereby implying $\sigma_{xy}(B)=-\sigma_{yx}(B)$
for the Hall conductivity of a reflection symmetric system.

Notwithstanding, noninteracting electrons constitute an exception to the above statement. For this case
it is sufficient to consider the most general Hamiltonian $H$ of a single particle, whose position
operator we denote by ${\bf r}_0$. In first quantization, ${\cal H}({\bf r})=\{H,\delta({\bf r}-{\bf r}_0)\}/2$,
where the curly brackets denote the anti-commutator.
Using the continuity equation, $-\bnabla\cdot{\bf J}^E=\partial_t{\cal H}=\{H,\partial_t \delta({\bf r}-{\bf r}_0)\}/2=
\bnabla\cdot\{H,{\bf J}^e\}/2e$, to identify the energy current density ${\bf J}^E$,
one finds for ${\bf J}^h={\bf J}^E+(\mu/e){\bf J}^e$
\be
\label{eq:jQni}
{\bf J}^h=-\frac{1}{2e}\{H-\mu,{\bf J}^e\}.
\ee
As a result, the correlation functions appearing in $\alpha_{ij}^{(1)}$ transform in the same way as the
$\langle {\rm J}^e_i {\rm J}^e_j\rangle$ correlation functions determining $\hat\sigma$. Specifically,
$\langle {\rm J}_i^e(t){\rm J}_j^h(0)\rangle_K=\langle {\rm J}_i^e(t)\{H-\mu,{\rm J}_j^e(0)\}\rangle_K/2e
=\langle \{H-\mu,{\rm J}_i^e(t)\}{\rm J}_j^e(t)\rangle_K/2e=\langle {\rm J}_i^h(t){\rm J}_j^e(0)\rangle_K$,
implying together with Eq. (\ref{eq:timerevsymmalpha}) that $\alpha_{ij}(B)=\alpha_{ji}(-B)$. This, in turn,
when combined with reflection symmetry, gives $\alpha_{xy}(B)=-\alpha_{yx}(B)$. However, we reiterate that
such a behavior is not guaranteed in the presence of interactions.


Let us note in passing that when $B=0$ the above discussion implies that for a generic interacting 
system with no reflection symmetry $\alpha_{ij}\neq \alpha_{ji}$ \cite{Hosur2013}. 
In this case it is impossible to make $\hat{\alpha}$ purely diagonal by choosing suitably 
aligned principle axes. Such an "anomalous" Peltier effect is different from the
Hall conductivity under the same conditions, which can always be made to 
vanish, and is necessarily a consequence of interactions, since in their absence 
$\alpha_{ij}=\alpha_{ji}$.

We now proceed to demonstrate the explicit violation of 
$\alpha_{xy}(B)=-\alpha_{yx}(B)$ in an exactly solvable example.

\subsection{Two interacting particles in a magnetic field}
\label{subsec:twopar}

Consider two interacting particles in a magnetic field, whose Hamiltonian
\ba
\label{eq:2pH}
\nonumber
\hspace{-1cm}H&=&H_0+U(\br_1-\br_2)=\frac{1}{2}\sum_{i=1,2}\left(m_x v_{i,x}^2+m_y v_{i,y}^2\right)\\
&&+\frac{1}{8}m_x\omega_x^2\left(x_1-x_2\right)^2+\frac{1}{8}m_y\omega_y^2\left(y_1-y_2\right)^2,
\ea
is reflection symmetric, but anisotropic due to the rotation asymmetry of the mass tensor and harmonic
interaction. The latter is characterized by the frequencies $\omega_{x,y}$, which together with the cyclotron
frequency, $\omega_c=eB/\sqrt{m_x m_y} c$, set the energy scales of the problem.
We work in the symmetric gauge for which the velocity operators take the form
\ba
\label{eq:velop}
v_x&=&\frac{1}{m_x}\left(p_x-\frac{eB}{2c}y\right), \\
v_y&=&\frac{1}{m_y}\left(p_y+\frac{eB}{2c}x\right).
\ea
The above Hamiltonian does not include a boundary potential, which is responsible for generating equilibrium
edge currents and magnetization. However, in a system much larger than the magnetic lengths
$l_{x,y}=1/\sqrt{m_{x,y}\omega_c}$ it has a negligible effect on the current
correlation functions in the bulk, which are our main point of interest.

Transforming to the center of mass coordinates, ${\bf R}=(\br_1+\br_2)/\sqrt{2}$, and relative coordinates
$\br=(\br_1-\br_2)/\sqrt{2}$, separates the Hamiltonian into two commuting sectors $H=H_{\rm CM}+H_r$,
with
\ba
\label{eq:Hsplit}
\nonumber
H_{\rm CM}&=&\frac{\omega_c}{2}\left[\left(-i l_x \frac{\partial}{\partial X}-\frac{Y}{2l_y}\right)^2
+\left(-i l_y \frac{\partial}{\partial Y}+\frac{X}{2l_x}\right)^2\right], \\
\nonumber
H_r&=&\frac{\omega_c}{2}\left[\left(-i l_x \frac{\partial}{\partial x}-\frac{y}{2l_y}\right)^2
+\left(-i l_y \frac{\partial}{\partial y}+\frac{x}{2l_x}\right)^2\right] \\
&&+\frac{1}{4\omega_c}\left[\omega_x^2\left(\frac{x}{l_x}\right)^2+\omega_y^2\left(\frac{y}{l_y}\right)^2\right].
\ea

Defining the complex coordinate $Z=X/l_x+iY/l_y$ and the operators
\ba
\label{eq:a1def}
a_1&=&\frac{Z^*}{2^{3/2}}+2^{1/2}\frac{\partial}{\partial Z},\\
\label{eq:a2def}
a_2&=&\frac{Z}{2^{3/2}}+2^{1/2}\frac{\partial}{\partial Z^*},
\ea
satisfying $[a_1,a_1^\dagger]=[a_2,a_2^\dagger]=1$ and $[a_1,a_2]=[a_1,a_2^\dagger]=0$, leads to the
familiar diagonalized form of $H_{\rm CM}$
\be
\label{eq:diagHCM}
H_{\rm CM}=\omega_c\left(a_1^\dagger a_1+\frac{1}{2}\right).
\ee

The relative Hamiltonian can be diagonalized via a series of canonical transformations that are
detailed in Appendix \ref{app:2pdiag}. The result is
\be
\label{eq:diagHrmain}
H_r=\omega_1\left(d_1^\dagger d_1+\frac{1}{2}\right)+\omega_2\left(d_2^\dagger d_2+\frac{1}{2}\right),
\ee
where $[d_1,d_1^\dagger]=[d_2,d_2^\dagger]=1$ and $[d_1,d_2]=[d_1,d_2^\dagger]=0$, and the frequencies
$\omega_{1,2}$ are given in Eq. (\ref{eq:omega12def}).
The energy eigenstates $|N,n\rangle\equiv |N_1,N_2,n_1,n_2\rangle$ are therefore characterized by the
eigenvalues of $a_1^\dagger a_1$, $a_2^\dagger a_2$, $d_1^\dagger d_1$ and $d_2^\dagger d_2$, respectively,
with energies $E_{N,n}=\omega_c\left(N_1+1/2\right)+\omega_1\left(n_1+1/2\right)+
\omega_2\left(n_2+1/2\right)$. The fermionic statistics forces odd $n_1+n_2$, see Appendix \ref{app:2pdiag}.

The first quantized form of Eq. (\ref{eq:Jedef}),
$\bJ^e(\br)=-(e/2)\sum_{i=1,2}\{{\bm v}_i,\delta_i\}$, where $\delta_i=\delta(\br-\br_i)$,
leads to the averaged electrical current density ${\bm J}^e=(1/A)\int d^2r \bJ^e(\br)$ with
\ba
\label{eq:Jex2p}
J^e_x&=&-i\frac{e\omega_c l_x}{A}(a_1^\dagger-a_1), \\
J^e_y&=&-\frac{e\omega_c l_y}{A}(a_1^\dagger+a_1).
\ea
An explicit calculation then readily confirms that the electrical current correlation functions
satisfy $\Tr\left[e^{-\beta H}J^e_x(t)J^e_y(0)\right]=
-\Tr\left[e^{-\beta H}J^e_y(t)J^e_x(0)\right]$, as required for $\sigma_{xy}=-\sigma_{yx}$.

It follows from the results of Appendix \ref{app:currents} that the averaged energy current density
takes the form
\ba
\label{eq:JEdef2p}
\nonumber
{\bm J}^E&=&\frac{1}{4A}\sum_{i=1,2}{\bm v}_i\left[m_x v_{i,x}^2+m_y v_{i,y}^2+U(\br_1-\br_2)\right]\\
&&+\frac{1}{4e}(\br_1-\br_2)\left[{\bm J}^e\cdot\frac{\partial U(\br_1-\br_2)}{\partial \br_1}\right] +{\rm H.c.},
\ea
where the commutativity of ${\bm J}^e$ with $\br_1-\br_2$ has been used. We are interested in the
correlation functions
\ba
\label{eq:JEcorfun2p}
\nonumber
&&\!\!\!\!\!\!\!\!\!\Tr\left[e^{-\beta H}J^e_x(t)J^E_y(0)\right]=\sum_{N,N',n}e^{(it-\beta)E_{N,n}-it E_{N',n}} \\
&&\hspace{2cm}\times\langle N,n|J^e_x|N',n\rangle\langle N',n|J^E_y|N,n\rangle,
\ea
and $\Tr\left[e^{-\beta H}J^e_y(t)J^E_x(0)\right]$, relevant to $\alpha_{xy}$ and $\alpha_{yx}$.
We therefore require only the piece in ${\bm J}^E$ which is
diagonal in $n_1,n_2$. Calculation reveals that the corresponding piece in $J^E_x$ may be expressed
as $\{I_x,J^e_x\}$, with
\ba
\label{eq:JExdiag}
\nonumber
I_x&=&\frac{\Omega}{8e}\left[\frac{\omega_x^2}{2\Omega^2}\cos^2\phi\, e^{-2\theta_1}
-\left(\cos\phi+\frac{\omega_c}{2\Omega}\sin\phi\right)^2e^{2\theta_1} \right] \\
\nonumber
&&\times \left(d_1^\dagger d_1+d_1 d_1^\dagger\right) \\
\nonumber
&+&\frac{\Omega}{8e}\left[\frac{\omega_x^2}{2\Omega^2}\sin^2\phi\, e^{2\theta_2}
-\left(\sin\phi-\frac{\omega_c}{2\Omega}\cos\phi\right)^2 e^{-2\theta_2}
\right] \\
&&\times\left(d_2^\dagger d_2+d_2 d_2^\dagger\right)-\frac{1}{4e}H,
\ea
where the parameters $\Omega$, $\phi$ and $\theta_{1,2}$ are given in Appendix \ref{app:2pdiag}.
At the same time the corresponding piece in $J^E_y$ reads $\{I_y,J^e_y\}$, with
\ba
\label{eq:JEydiag}
\nonumber
I_y&=&\frac{\Omega}{8e}\left[\frac{\omega_y^2}{2\Omega^2}\sin^2\phi\, e^{2\theta_1}
-\left(\sin\phi+\frac{\omega_c}{2\Omega}\cos\phi\right)^2e^{-2\theta_1}\right] \\
\nonumber
&&\times \left(d_1^\dagger d_1+d_1 d_1^\dagger\right)\\
\nonumber
&+&\frac{\Omega}{8e}\left[\frac{\omega_y^2}{2\Omega^2}\cos^2\phi\, e^{-2\theta_2}
-\left(\cos\phi-\frac{\omega_c}{2\Omega}\sin\phi\right)^2 e^{2\theta_2}\right] \\
&&\times\left(d_2^\dagger d_2+d_2 d_2^\dagger\right)-\frac{1}{4e}H.
\ea
Since $[I_x,H]=[I_y,H]=0$, the same argument presented following Eq. (\ref{eq:jQni}) would
imply that $\Tr\left[e^{-\beta H}J^e_x(t)J^E_y(0)\right]=
-\Tr\left[e^{-\beta H}J^e_y(t)J^E_x(0)\right]$, provided that $I_x=I_y$.
However, this condition is fulfilled only when $\omega_x=\omega_y$,
leading to $\cos\phi=\pm\sin\phi=1/\sqrt{2}$ and $\theta_1=\theta_2=0$.
Hence, we conclude that $\alpha_{xy}\neq-\alpha_{yx}$ except when the system is
isotropic ($m_x=m_y$ and $\omega_x=\omega_y$), or when the anisotropy in the interaction
matches the mass anisotropy ($m_x\neq m_y$ and $\omega_x=\omega_y$), in which case it may
be removed by coordinate rescaling.

\vspace{-0.5cm}
\begin{figure}[t]
\centering
  \includegraphics[width=\linewidth,clip=true]{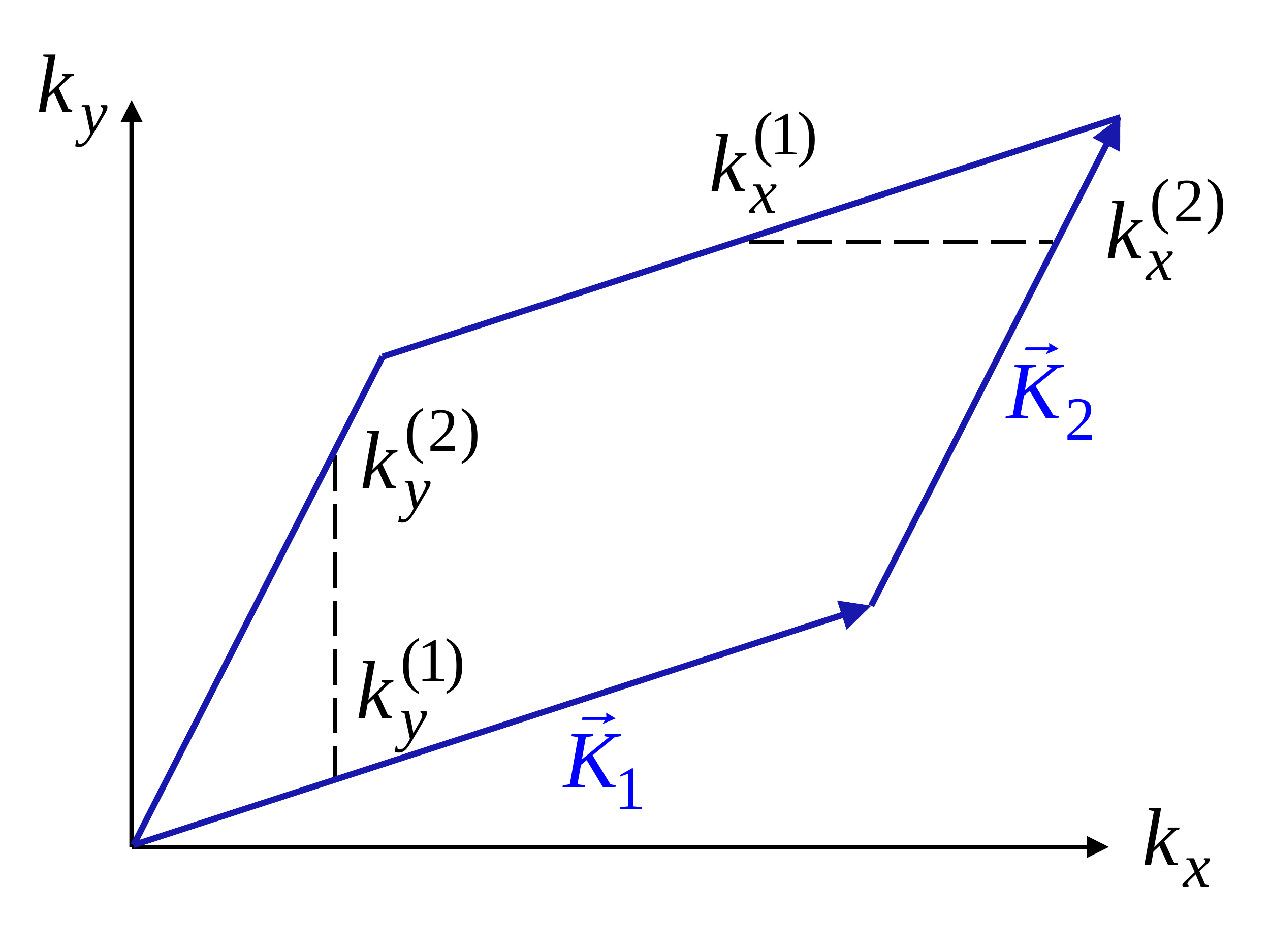}
  \caption{The integration region in $\bk$ space.}
  \label{fig:BO}
\end{figure}

\subsection{{\large $\hat\alpha$} within Boltzmann transport theory}
\label{subsec:Boltzmann}

Let us next apply the Boltzmann equation to the transport of spinless electrons in a two-dimensional system
subjected to a perpendicular magnetic field ${\bf B}=B\hat{\bf z}$. This approach is appropriate on time
and length scales much larger than the corresponding scales characterizing the scattering events. Consequently,
the effects of scattering are captured by a local collision integral. Close to
equilibrium, the distribution function can be written as $f_\bk-(\partial f_\bk/\partial\varepsilon_\bk) g_\bk$,
with $f_\bk=n_F(\varepsilon_\bk)$ and $\beta g_\bk\ll f_\bk$. As a result, the collision integral takes the form
$-\int_{\bk'} I_{\bk,\bk'} g_{\bk'}$, where the kernel $I_{\bk,\bk'}=I_{\bk',\bk}$ depends on the
equilibrium transition rates \cite{Ziman,Abrikosov}, and the integral $\int_\bk\equiv\int d^2k/(2\pi)^2$
extends over the reciprocal unit cell spanned by the vectors ${\bf K}_{1,2}$, see Fig. \ref{fig:BO} .
To linear order in the applied homogeneous electric field and thermal gradient the Boltzmann equation
reads \cite{Ziman,Abrikosov}
\be
\label{eq:linBol}
B_\bk g_\bk +\int_{\bk'} I_{\bk,\bk'}g_{\bk'}
=\bv_\bk\cdot\left[e\bE+\left(\varepsilon_\bk-\mu\right)\frac{\bnabla T}{T}\right]
\frac{\partial f_\bk}{\partial\varepsilon_\bk},
\ee
where we have assumed that the energy spectrum consists of a single band and defined the differential operator
\ba
\label{eq:Bk}
\nonumber
B_\bk&=&-\frac{e}{\hbar c}\frac{\partial f_\bk}{\partial\varepsilon_\bk}
\left(\bv_\bk\times{\bf B}\right)\cdot\bnabla_\bk \\
&=&\frac{eB}{\hbar^2 c}\frac{\partial f_\bk}{\partial\varepsilon_\bk}
\left(\frac{\partial\varepsilon_\bk}{\partial k_x}\frac{\partial}{\partial k_y}-
\frac{\partial\varepsilon_\bk}{\partial k_y}\frac{\partial}{\partial k_x}\right).
\ea


Solving Eq. (\ref{eq:linBol}) yields
\ba
\label{eq:solgk}
\nonumber
g_\bk&=&\int_{\bk_0}I_{\bk,\bk_0}^{-1}\bv_{\bk_0}\cdot\left[e\bE+\left(\varepsilon_{\bk_0}-\mu\right)
\frac{\bnabla T}{T}\right]\frac{\partial f_{\bk_0}}{\partial\varepsilon_{\bk_0}} \\
\nonumber
&+&\int_{\bk_0}I_{\bk,\bk_0}^{-1}\sum_{n=1}^\infty (-1)^n\prod_{m=1}^n \int_{\bk_m}B_{\bk_{m-1}}I_{\bk_{m-1},\bk_m}^{-1} \\
&&\times \bv_{\bk_n}\cdot\left[e\bE+\left(\varepsilon_{\bk_n}-\mu\right)
\frac{\bnabla T}{T}\right]\frac{\partial f_{\bk_n}}{\partial\varepsilon_{\bk_n}},
\ea
where $\int_{\bk'} I_{\bk,\bk'}I_{\bk',\bk''}^{-1}=(2\pi)^2\delta(\bk-\bk'')$.
Since the electrical and heat current densities are given by ${\bm J}^e=-e \int _\bk \bv_\bk \delta f_\bk$,
and ${\bm J}^h=\int _\bk \bv_\bk (\varepsilon_\bk-\mu)\delta f_\bk$
it follows that
\ba
\label{eq:alphabol1}
\nonumber
\left( \begin{array}{c}
T\alpha_{ij}\\
\tilde\alpha_{ij} \\ \end{array}\right)&=&e\int_\bk\int_{\bk_0}
\frac{\partial f_{\bk}}{\partial\varepsilon_{\bk}}
v_{i,\bk}I_{\bk,\bk_0}^{-1}v_{j,\bk_0}
\frac{\partial f_{\bk_0}}{\partial\varepsilon_{\bk_0}}\left( \begin{array}{c}
\varepsilon_{\bk_0}-\mu\\
\varepsilon_\bk-\mu \\ \end{array}\right) \\
\nonumber
&+&e\int_\bk\int_{\bk_0}
\frac{\partial f_{\bk}}{\partial\varepsilon_{\bk}}
v_{i,\bk}I_{\bk,\bk_0}^{-1}\sum_{n=1}^\infty(-1)^n \prod_{m=1}^n\int_{\bk_m} \\
&\times& B_{\bk_{m-1}}I_{\bk_{m-1},\bk_m}^{-1} v_{j,\bk_n}
\frac{\partial f_{\bk_n}}{\partial\varepsilon_{\bk_n}}\left( \begin{array}{c}
\varepsilon_{\bk_n}-\mu\\
\varepsilon_\bk-\mu \\ \end{array}\right).
\ea

The above result obeys the Onsager relation (\ref{eq:timerevsymmalpha}) at  $B=0$, as can be readily
verified by using the symmetry $I_{\bk,\bk'}=I_{\bk',\bk}$ and exchanging $\bk \leftrightarrow \bk_0$
in the first line of Eq. (\ref{eq:alphabol1}). To demonstrate that the Onsager relation continues to
hold for $B>0$ we integrate by parts the integrals in the second line, use the symmetry of the collision
kernel and exchange $\bk \leftrightarrow \bk_n$, $\bk_m \leftrightarrow \bk_{n-m-1}$ for
$m=0,\cdots, \lceil (n-1)/2\rceil$. This brings the expression back to itself up to $B_\bk\rightarrow -B_\bk$,
and $\varepsilon_\bk \leftrightarrow \varepsilon_{\bk_n}$ in the last parenthesis. Accordingly, the desired
relation is established, provided that the contribution from
the boundary terms, incurred during the integration by part, vanishes.
On general grounds, $\varepsilon_{\bk+{\bf K}}=\varepsilon_\bk$ and
$\bv_{\bk+{\bf K}}=\bv_\bk=(1/\hbar)\partial\varepsilon_\bk/\partial\bk$, for any reciprocal vector ${\bf K}$.
We find that the boundary contribution vanishes if $I_{\bk,\bk'}$ also respects the lattice periodicity,
\ie, $I_{\bk+{\bf K},\bk'}=I_{\bk,\bk'}$. Under such conditions the integrand is invariant under translation
by a reciprocal wave-vector and for every contribution from an end point
$[k_x,k_y^{(1)}(k_x)]$ there exists an opposite contribution from an end point at $[k_x,k_y^{(1)}(k_x)]+{\bf K}_2$
or $[k_x,k_y^{(1)}(k_x)]-{\bf K}_1$, see Fig. \ref{fig:BO}. A similar argument works for the other end points.

The preceding analysis shows that $\alpha_{ij}(B)= \alpha_{ji}(-B)$, and therefore $\alpha_{ij}(B)=-\alpha_{ji}(B)$
in reflection symmetric systems, only if $\varepsilon_\bk =\varepsilon_{\bk_n}$ in Eq. (\ref{eq:alphabol1}).
This condition is fulfilled whenever $I_{\bk,\bk'}^{-1}$ is proportional to
$\delta(\varepsilon_\bk-\varepsilon_{\bk'})$, as is the case for elastic impurity scattering, or
within the relaxation time approximation where
$I_{\bk,\bk'}^{-1}=\delta(\bk-\bk')(\partial f_\bk/\partial\epsilon_\bk)^{-1}\tau_\bk$.
An important case of interest is the disordered Fermi liquid which includes both elastic impurity scattering
and inelastic processes due to electron-electron interactions. While the elastic piece in $I_{\bk,\bk'}^{-1}$
is temperature independent, the inelastic channel contribution to $I_{\bk,\bk'}^{-1}$ scales as $T^2$ in three
dimensions \cite{Ziman,Abrikosov}. Therefore, at low temperatures the physics is dominated by the former,
$\alpha_{xy}\sim T$, and the relation $\alpha_{ij}(B)=-\alpha_{ji}(B)$ holds up to corrections of order $T^3$.
In the following section we turn our attention to the behavior of $\hat\alpha$ in a system which is manifestly
a non-Fermi liquid.

\section{The Nernst effect in a system of coupled Luttinger liquids}

\label{sec:Nernst}

\subsection{The model and its {\normalsize $\hat\alpha$}}
\label{subsec:Nernst-alpha}

We consider a model of a two-dimensional array of $N_c$ one-dimensional chains
extending along the $x$ direction from $-L/2$ to $L/2$ and separated by a distance $d$ in the
$y$ direction, with both $N_c,L\rightarrow\infty$. The chains are immersed in a magnetic field
${\bf B}=B\hat{\bf z}$, which is generated by the vector potential $A_y=Bx$. The spinfull electrons
that populate the system interact via an attractive contact interaction, which opens a gap in the
spin sector of each chain \cite{dimcross}. This gap is assumed to be much larger than
any remaining energy scale in the problem, such as the temperature and inter-chain couplings.
Owing to the spin gap, single-particle tunneling between the chains is irrelevant. In contrast,
the superconducting and $2k_F$ charge-density wave (CDW) susceptibilities of the chains are enhanced
and the inter-chain Josephson and CDW couplings are important \cite{dimcross}. In order to have a
non-trivial transverse thermoelectric response one needs to include the Josephson tunneling.
We will neglect the CDW coupling, whose main effect is to compete against the superconducting
ordering tendency of the system, since we are interested in the case where the latter prevails.
Consequently, we study the following bosonized form of $H=H_0+H_{\sJ}$, where
\ba
\label{eq:1dH0}
\!\!\!\!\!\!\!\!\!H_0&=&\frac{v}{2}\sum_{j=1}^{N_c}\int dx \left[ K\left(\partial_x\theta_j\right)^2+\frac{1}{K}
\left(\partial_x\phi_j\right)^2\right], \\
\label{eq:1dHJ}
\!\!\!\!\!\!\!\!\!H_{\sJ}&=&-\J\sum_{j=2}^{N_c}\int dx \cos\left[\sqrt{2\pi}\left(\theta_{j}-\theta_{j-1}\right)+bx\right].
\ea
Here $v$ and $K>1$ are the velocity and Luttinger parameter of the charge sector, respectively. $\J$ is the
Josephson energy per unit length, and $b=2eBd/c=2d/l_B^2$ is a wavevector associated with the magnetic field.
Eq. (\ref{eq:1dHJ}) shows that the field adds an oscillatory phase to the pair hopping term, thus rendering it
irrelevant in the renormalization group sense. However, a second order term in $\J$ is relevant for $K>3/2$
and induces a crossover to a strong coupling regime at $T_c\sim (v/a)(\J/v)^{K/(K-3/2)}$, where $a$ is the
short distance cutoff of the theory \cite{Sondhi-sliding}.
Therefore, the perturbative treatment of $\J$, which we employ below, is valid only for $T>T_c$.

The current density operators may be deduced from the continuity equations for the conserved
quantities. For the average current densities we obtain
\ba
\label{eq:Jex1d}
\!\!\!\!\!\!\!\!\!J^e_x&=&-\sqrt{\frac{2}{\pi}}\frac{evK}{A}\sum_{j=1}^{N_c}\int dx\partial_x\theta_j, \\
\label{eq:Jey1d}
\!\!\!\!\!\!\!\!\!J^e_y&=&-\frac{2e\J d}{A}\sum_{j=2}^{N_c}\int dx \sin\left[\sqrt{2\pi}\left(\theta_j
-\theta_{j-1}\right)+bx\right], \\
\label{eq:Jhx1d}
\!\!\!\!\!\!\!\!\!J^h_x&=&-\frac{v^2}{2A}\sum_{j=1}^{N_c}\int dx\left\{\partial_x\phi_j,\partial_x\theta_j\right\}, \\
\label{eq:Jhy1d}
\nonumber
\!\!\!\!\!\!\!\!\!J^h_y&=&-\frac{\sqrt{2\pi}v\J d}{4KA}\sum_{j=2}^{N_c}\int dx \\
\!\!\!\!\!\!\!\!\!&\times&\left\{\partial_x\phi_j+\partial_x\phi_{j-1},
\sin\left[\sqrt{2\pi}\left(\theta_j-\theta_{j-1}\right)+bx\right]\right\},
\ea
where $A=LN_cd$. Note that in the Luttiner model (\ref{eq:1dH0}) the energy is measured
relative to the chemical potential and therefore $\bJ^h$ is calculated from the
continuity equation for the Hamiltonian density.

Using the above expressions and Eq. (\ref{eq:alpha1}) we compute $\alpha^{(1)}_{yx}$ to second order in $\J$,
see Appendix \ref{app:1dalpha} for details. The result
\be
\label{eq:alpha1yx1d}
\alpha_{yx}^{(1)}=\lim_{\omega\rightarrow 0}-\frac{ebv^2\J^2}{2T}\frac{\partial C(b,\omega)}{\partial\omega^2},
\ee
is expressed in terms of the function
\ba
\label{eq:Cqomega}
\nonumber
C(q,\omega)&=&\frac{a^2}{v}\sin\left(\frac{\pi}{K}\right)\left(\frac{l_T}{2a}\right)^{2-2/K}\\
\nonumber
&&\times B\left[\frac{1}{2K}-\frac{i}{4}\left(\frac{\omega}{v}-q\right)l_T,1-\frac{1}{K}\right]\\
&&\times B\left[\frac{1}{2K}-\frac{i}{4}\left(\frac{\omega}{v}+q\right)l_T,1-\frac{1}{K}\right],
\ea
where $B(x,y)$ is the beta function, and $l_T=v/\pi T$ is the thermal length. Appendix \ref{app:1dalpha}
also contains the computation of $M_z$, which, together with Eq. (\ref{eq:alpha2}), leads to
\be
\label{eq:alpha2yx1d}
\alpha_{yx}^{(2)}=-\frac{e\J^2}{2T}\frac{\partial C(b,0)}{\partial b}.
\ee

\begin{figure}[t]
\centering
  \includegraphics[width=\linewidth,clip=true]{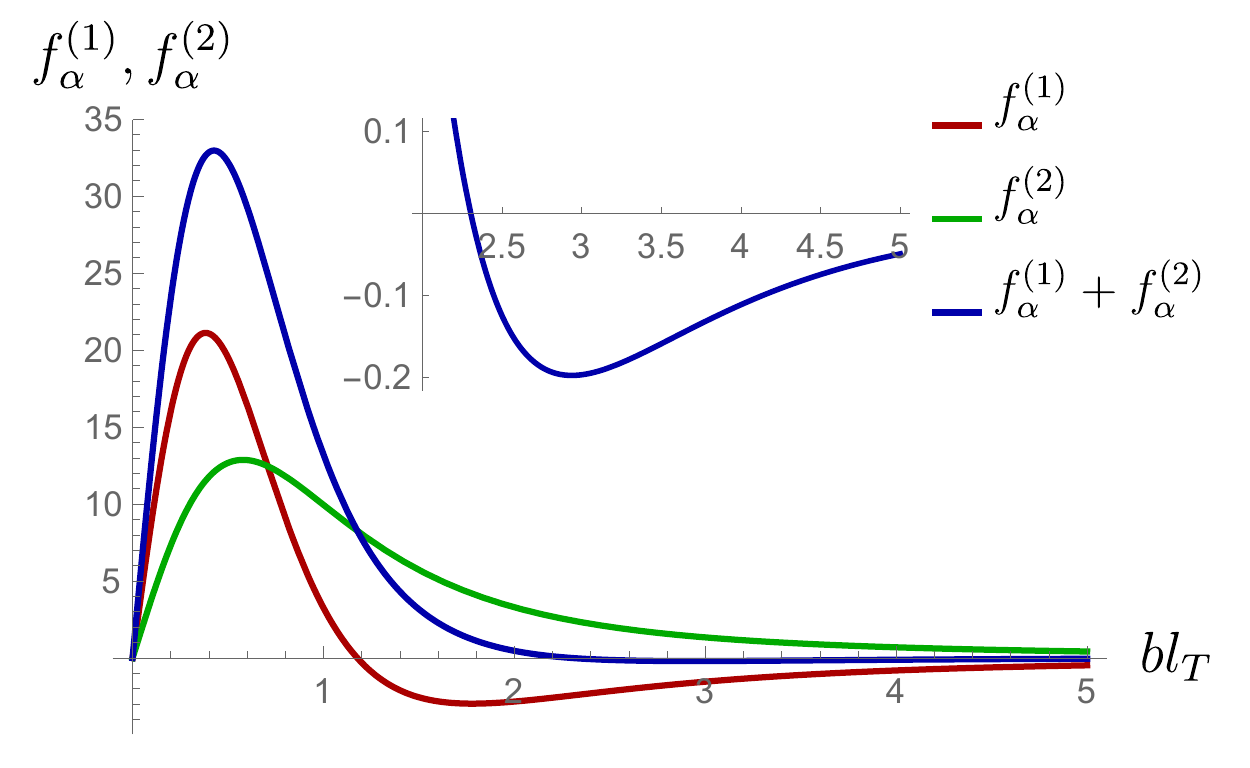}
  \caption{The scaling functions that determine $\alpha_{yx}$, shown here for
$K=2$. The inset depicts the sign change of $f_\alpha^{(1)}+f_\alpha^{(2)}$, and thus of
$\alpha_{yx}$ for large $bl_T$.}
  \label{fig:alphascaling}
\end{figure}

The final result for $\alpha_{yx}$ may be cast into a scaling form
\be
\label{eq:f1scaling}
\alpha_{yx}=e\left(\frac{\J a^2}{v}\right)^2\left(\frac{l_T}{a}\right)^{4-2/K}\left[f_\alpha^{(1)}(bl_T)
+f_\alpha^{(2)}(bl_T)\right],
\ee
where the functions $f_\alpha^{(1,2)}$ originate from $\alpha_{yx}^{(1,2)}$, respectively. Both $f_\alpha^{(1)}(x)$ and
$f_\alpha^{(2)}(x)$ scale as $x$ for $x\ll 1$, and decay as $x^{-(3-2/K)}$ for $x\gg 1$, due to the rapid oscillations
in the Josephson coupling, see Fig. \ref{fig:alphascaling}. While $\alpha_{yx}^{(2)}$ is always positive, consistent
with a diamagnetic response ($M_z<0$), the sign of $\alpha_{yx}^{(1)}$ changes as function of $bl_T$. At weak fields
and high temperatures the two contributions add up, leading to a positive $\alpha_{yx}$, which behaves as $B/T^{5-2/K}$.
On the other hand, at large magnetic fields and low temperatures they tend to cancel each other leaving a total negative
$\alpha_{yx}$, which varies according to $-T/B^{5-2/K}$. The sign of $\alpha_{yx}$ in this regime is the one expected
from superconducting fluctuations.

In contrast, we show in Appendix \ref{app:1dalpha} that $\alpha^{(1)}_{xy}$ is smaller by a factor $l_T/L$ than
$\alpha^{(1)}_{yx}$, and hence negligible in the thermodynamic limit. This is a consequence of the fact that
in the clean model considered here $[J^e_x,H]=0$, up to corrections from boundary terms. As a result
the retarded $J^e_xJ^h_y$ correlation function which determine $\alpha^{(1)}_{xy}$ vanishes identically.
This demonstrates that in the inherently interacting problem studied here, $\alpha_{xy}=cM_z/AT\neq-\alpha_{yx}$.
We expect that upon breaking the conservation of $J^e_x$, \eg, by introducing disorder into the chains,
$\alpha^{(1)}_{xy}$ will no longer vanish. Nevertheless, its magnitude will be proportional to the disorder
strength and will not match that of $\alpha^{(1)}_{yx}$.

Let us comment that a model for two
superconducting wires, similar to $H_0+H_{\sJ}$ defined by Eqs. (\ref{eq:1dH0}) and (\ref{eq:1dHJ}), was considered
in Ref. \onlinecite{Efrat-Nernst}. However, unlike the present study each wire was assumed to be in equilibrium,
described by a density matrix $e^{-H_0/T}$ with a different temperature, while the Josephson coupling was turned on
adiabatically. Consequently, it was found that $\alpha_{xy}=0$. Upon including a term which breaks the linear dispersion
and characterized by a dimensionless curvature ${\cal C}$, this result changed to $\alpha_{xy}=-cM_z/AT_0$, where
$T_0=v/(\pi a{\cal C})$.

\subsection{The conductivity and Nernst signal}
\label{subsec:Nernst-signal}

For a particle-hole and reflection symmetric model, such as the one considered here, the relation between
the Peltier coefficients and the thermopower is considerably simplified. Under particle-hole
transformation ${\bm J}^e(B)\rightarrow -{\bm J}^e(-B)$ and ${\bm J}^h(B)\rightarrow {\bm J}^h(-B)$. Therefore, in
the symmetric case, where $K(B)\rightarrow K(-B)$, we conclude that $\hat\sigma(B)=\hat\sigma(-B)$
and $\hat\alpha(B)=-\hat\alpha(-B)$. When combined with Eq. (\ref{eq:refsymmalpha}) due to reflection
symmetry, it leads to the result $\sigma_{xy}=\sigma_{yx}=\alpha_{xx}=\alpha_{yy}=0$. In turn, one finds
for the Nernst signals
\be
\label{eq:Nernstsignals}
S_{xy}=\frac{\alpha_{xy}}{\sigma_{xx}}, \;\;\;\;\; -S_{yx}=-\frac{\alpha_{yx}}{\sigma_{yy}}.
\ee

For a quasi-one-dimensional system embedded in a magnetic field and possessing Galilean invariance along the chains
one finds $\sigma_{xx}\sim 1/\kappa^2$, where $\kappa$ is the curvature of the free chain spectrum \cite{Giamarchi-Hall}.
In our linearized model $\sigma_{xx}$ diverges and as a result $S_{xy}=0$.
To calculate $\sigma_{yy}$ we apply
Eq. (\ref{eq:PiChi}) (with $\bJ=\bJ^e$) and find to second order in $\J$
\ba
\label{eq:sigmayy}
\nonumber
\sigma_{yy}&=&\lim_{\omega\rightarrow 0}-2ide^2\J^2\frac{\partial C(b,\omega)}{\partial\omega} \\
&=&e^2\frac{d}{a}\left(\frac{\J a^2}{v}\right)^2\left(\frac{l_T}{a}\right)^{3-2/K}f_\sigma(bl_T).
\ea

\begin{figure}[t]
\centering
  \includegraphics[width=\linewidth,clip=true]{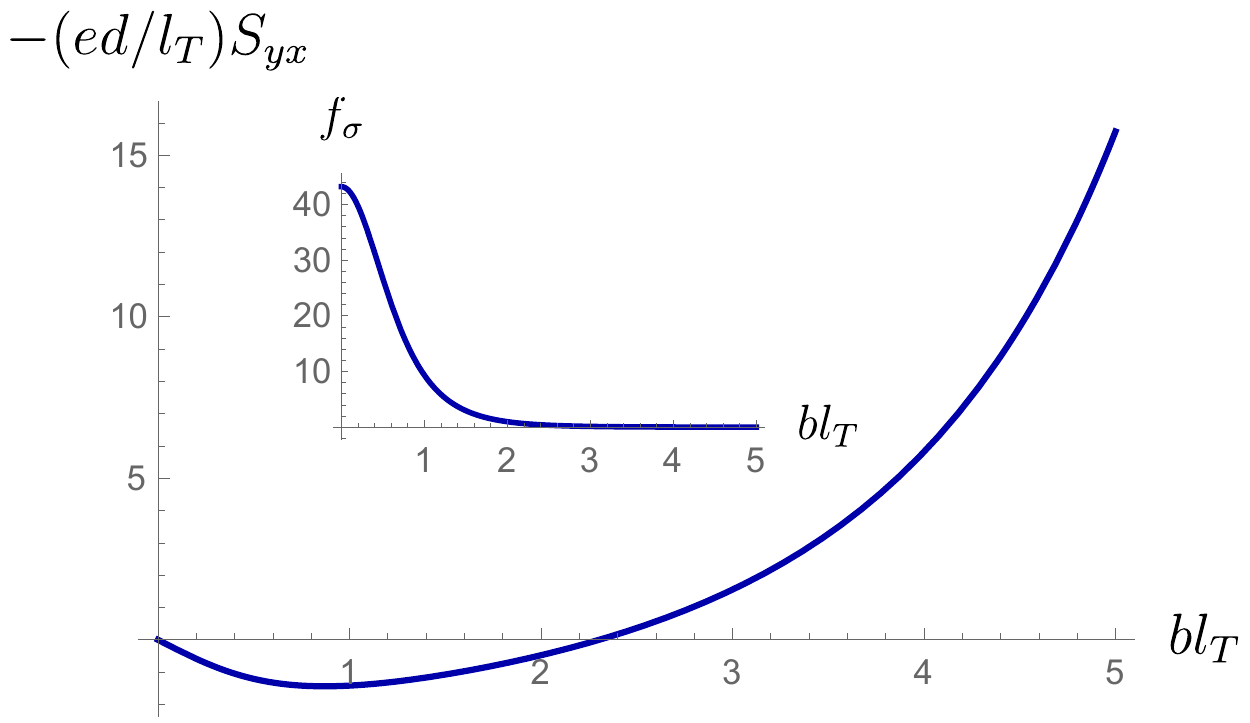}
  \caption{The dimensionless Nernst signal $-(ed/l_T)S_{yx}$ for the case
$K=2$. The inset depicts the scaling function of the $\sigma_{yy}$.}
  \label{fig:Nernst}
\end{figure}

The conductivity scaling function $f_\sigma(x)$ is depicted in the inset of Fig, \ref{fig:Nernst}. It
tends to a constant at small $x$ and decays as $x^{-(2-2/K)}e^{-\pi x/2}$ for large $x$.
When combined with the behavior of $\alpha_{yx}$ this results in a Nernst signal along the $y$ direction
that is negative and scales according to $B/T^2$ for low fields and large temperatures ($bl_T\ll 1$). As the
field is increased and the temperature lowered the Nernst signal turns positive and eventually, when $bl_T\gg 1$,
follows $(T^2/B^3)e^{(evd/c)(B/T)}$, see Fig, \ref{fig:Nernst}. The resulting scale for the Nernst signal, $l_T/ed$,
is very large. For typical values relevant for the quasi-one-dimensional conductors, $v=10^5$ ms$^{-1}$, $d=1$ nm,
and $T=10$  K, the Nernst signal is of order $S_{yx}\approx 1$ mV K$^{-1}$, to be compared with values of order
0.1 mV K$^{-1}$, measured in (TMTSF)$_2$ClO$_4$ \cite{1d-Nam}. The Nernst coefficient, $e_N=-S_{yx}/B$, calculated
for low fields where $S_{yx}$ is linear in $B$, is also large. For the above parameters we find
$e_N\approx 100$ $\mu$V K$^{-1}$ T$^{-1}$, while $e_N$ measured in (TMTSF)$_2$ClO$_4$
is of order 10 $\mu$V K$^{-1}$ T$^{-1}$ \cite{1d-Nam}. This is in contrast to $e_N\approx 0.1$ $\mu$V K$^{-1}$ T$^{-1}$
calculated using Boltzmann theory for a similar band structure \cite{1d-Nam}.

\section{Conclusions}

\label{sec:Conclusions}

The transformation properties of a system under spatial reflections, time reversal and charge conjugation
relate many of its transport coefficients. Here we showed that the frequently used relation
$\alpha_{xy}=-\alpha_{yx}$ does not belong to this category. Rather, its validity requires the
additional condition of no interactions between the electrons, either directly or via mediators such
as phonons. Nevertheless, it becomes a good approximation whenever the interacting system can be
considered to comprise of weakly and locally interacting particles, \ie, a Fermi liquid.
In that sense, the above relation is similar to the Wiedemann-Franz law. They both reflect an
underlying assumption that heat transfer is restricted to convection by motion of the
charge carriers. The violation of $\alpha_{xy}=-\alpha_{yx}$ in a reflection symmetric system is therefore
a clear sign that energy is also transported via interactions between the particles, or in the extreme limit
that the concept of a quasiparticle fails. Thus, it would be interesting to follow the relation between
$\alpha_{xy}$ and $\alpha_{yx}$ as function of temperature. If, for example,
$r=(\alpha_{xy}+\alpha_{yx})/(\alpha_{xy}-\alpha_{yx})\ll 1$ is observed at high temperatures but
approaches $r\approx 1$ below a characteristic temperature, $T_0$, this would
mean one of the following: (i) $T_0$ indicates a nematic transition inside a non-Fermi liquid state, \ie,
a breaking of the $C_4$ rotation symmetry around the $z$ axis. (ii) The system is anisotropic and breaks
the reflection symmetry about the $x$ and $y$ directions below $T_0$. (iii) The system is a Fermi-liquid
and breaks both reflection symmetry and $C_4$ rotation symmetry at low temperatures. (iv) The system
is anisotropic but reflection symmetric and non-Fermi liquid behavior onsets at the temperature scale $T_0$.
The pseudogap regime of the high-temperature superconductors, with its tendencies to develop various
ordered states, seems to be a good candidate for such an experiment.

By studying the Nernst effect in an interacting quasi-one-dimensional model with strong superconducting
fluctuations we were able to demonstrate that the effect is much stronger than in two-dimensional models
considered using Boltzmann transport theory. This finding points to the importance of interactions and
low dimensionality in establishing a large Nernst signal, and may bare relevance to experiments done
on quasi-one-dimensional materials.

\acknowledgments
We would like to thank Steve Kivelson, Eun-Ah Kim and Kamran Behnia for helpful discussions. 
This research was supported by the United States-Israel Binational Science Foundation
(Grant No. 2014265) and by the Israel Science Foundation (Grant No. 585/13).

\appendix
\section{The current density operators}
\label{app:currents}

Here we obtain the electrical and heat current density operators of model (\ref{eq:2dhdensity}).
To begin with, the continuity equation for the charge density $\rho^e=-e\rho$ in the presence of a
gravitational field
\be
\label{eq:contcharge}
\partial_\mu {\rm J}^e_\mu=i\int d^2r' \left[1+g(\br')\right]\left[\rho^e(\br),\H(\br')\right],
\ee
is satisfied by the electrical current density operator
\be
\label{eq:Jedef}
{\rm J}^e_j(\br)=\left[1+g(\br)\right]\frac{ie}{2m_j}\psi^\dagger(\br)D_j\psi(\br)+\rm{H.c.}.
\ee

The heat current density operator $\bJ^h=\bJ^E+(\mu/e)\bJ^e$ is related to the energy current density
operator $\bJ^E$, which in turn is to be determined by the continuity equation for the energy density
\begin{widetext}
\ba
\label{eq:contenergy}
\nonumber
\partial_\mu {\rm J}^E_\mu=i\left[\H(\br),H \right]
&=&\partial_\mu\left\{\frac{i}{2m_\mu}\left[D_\mu\psi(\br)\right]^\dagger
\left[-\frac{1}{2m_\nu}D_\nu^2-e\phi(\br)+\frac{1}{2}\int d^2r' U(\br-\br')\rho(\br')\right]\psi(\br)\right\}\\
&&+\frac{i}{4m_\mu}\int d^2r'\left\{\left[D_\mu\psi(\br)\right]^\dagger\left[\partial_\mu U(\br-\br')\right]
\rho(\br')\psi(\br)+\left(\br\leftrightarrow\br'\right)\right\}+{\rm H.c.}.
\ea
\end{widetext}
Here, in order to avoid a surface term which arises in the derivation, we have assumed that no
charge current is flowing out of the system, \ie, $\bJ^e\cdot \bn=0$, with $\bn$ the normal to
the system's boundary.

To make progress we need to integrate Eq. (\ref{eq:contenergy}) with the appropriate boundary conditions.
To this end, we assume that the system is thermally isolated in the sense $\bJ^E\cdot \bn=0$. Both conditions
on the currents are fulfilled if ${\bf D}\psi\cdot \bn=0$. Denoting the second line in Eq. (\ref{eq:contenergy})
by $F(\br)$ we further assume that its contribution to $\bJ^E$ is irrotational and hence can be expressed
as $\bnabla\Phi$, where $\nabla^2\Phi(\br)=F(\br)$. It follows from the divergence theorem that a solution
to this equation exists only if $\int d^2r F(\br)=0$, which holds true in our case. Consequently, we find
\begin{widetext}
\ba
\label{eq:JEdef}
\nonumber
{\rm J}^E_j(\br)&=&\frac{i}{2m_j}\left[D_j\psi(\br)\right]^\dagger
\left[-\frac{1}{2m_\mu}D_\mu^2-e\phi(\br)+\frac{1}{2}\int d^2r' U(\br-\br')\rho(\br')\right]\psi(\br)\\
&&+\frac{i}{4m_\mu}\int d^2r' d^2r'' \partial_j\left[G(\br,\br')-G(\br,\br'')\right]
\left[D'_\mu\psi(\br')\right]^\dagger \left[\partial'_\mu U(\br'-\br'')\right]
\rho(\br'')\psi(\br')+{\rm H.c.},
\ea
\end{widetext}
where $G(\br,\br')$ is the Green's function of the Laplace equation
with Neumannn boundary conditions, satisfying $\nabla^2 G(\br,\br')=\delta(\br-\br')-1/A$
and $\nabla G(\br,\br')\cdot \bn=0$. For a rectangular domain $A=L_x\times L_y$ it is given by
\be
G(\br,\br')=\sideset{}{'}\sum_{m,n=0}^\infty\frac{u_{mn}(\br)u_{mn}(\br')}{\lambda_{mn}},
\ee
where the term $m=n=0$ is excluded from the sum and the Laplacian eigenfuncions and
eigenvalues are given by
\ba
u_{mn}(\br)&=&\frac{c_{mn}}{\sqrt{A}}\cos\left(\frac{m\pi x}{L_x}\right)\cos\left(\frac{n\pi y}{L_y}\right),\\
\lambda_{mn}&=&-\left(\frac{m\pi}{L_x}\right)^2-\left(\frac{n\pi}{L_y}\right)^2,
\ea
with $c_{mn}=2^{[{\rm sign}(m)+{\rm sign}(n)]/2}$. Subsequently, it follows from
\be
\label{eq:intG}
\int d^2r \partial_j G(\br,\br')=\frac{L_j}{2}-r'_j,
\ee
that the average current density
\be
\label{eq:aveJop}
{\bm J}^E=\frac{1}{A}\int d^2r \bJ^E(\br),
\ee
is readily obtained from Eq. (\ref{eq:JEdef}) (up to the factor $1/A$) by integrating the first line
over $\br$ and replacing $\partial_j\left[G(\br,\br')-G(\br,\br'')\right]$ with $r''_j-r'_j$ in the second.
Alternatively, it can also be expressed as
\ba
\label{eq:intJE}
\nonumber
J^E_j&=&-\frac{1}{2m_jA}\int d^2r \left[D_j\psi(\br)\right]^\dagger\partial_t\psi(\br) \\
\nonumber
&&-\frac{1}{4A}\int d^2r d^2r'(r_j-r'_j)U(\br-\br') \\
\nonumber
&&\times\left\{\psi^\dagger(\br)\rho(\br')\partial_t\psi(\br)
-\psi^\dagger(\br)\psi^\dagger(\br')\left[\partial_t\psi(\br')\right]\psi(\br)\right\} \\
&& +{\rm H.c.}.
\ea

\section{The Kubo formula for the thermoelectric coefficients}
\label{app:Kubo}

Consider a time independent $K$ (with $[H,N]=0$), perturbed by $\delta K=\int d^2r g(\br,t) Q(\br)$,
where $g(\br,t)=g(\br)e^{-i(\omega+i\delta)t}$ is an external field coupled to a conserved charge $Q$,
satisfying $\partial_t Q+\bnabla \cdot\bJ=0$. To linear order in $g$ an observable $O(t)=e^{iKt}Oe^{-iKt}$,
with $\langle O(t)\rangle_{K}=0$, acquires the expectation value \cite{Mahan}
\be
\label{eq:Kuboeq1}
\langle O(t)\rangle_{K+\delta K}=\langle \delta O(t)\rangle_{K}-\int d^2r \bnabla \varphi(\br,t)\cdot
\Pi_{\bJ,O}(\br,\omega+i\delta).
\ee
Here we have assumed that no $\bJ$ flows out of the system. \ie, $\bJ\cdot \bn=0$, denoted by
$\delta O$ the change in the form of $O$ in the presence of $\delta K$, and
\be
\label{eq:Pi}
\Pi_{\bJ,O}(\br,\omega)=\int_0^\infty dt\int _0^\beta d\tau e^{i\omega t}\langle \bJ(\br,-t-i\tau)O(0)\rangle_K.
\ee
Using a Lehmann representation in terms of $K$ eigenstates, $K|n\rangle=\xi_n|n\rangle$,
we can write the latter as
\ba
\label{eq:PiLeh}
\nonumber
\Pi_{\bJ,O}(\br,\omega+i\delta)&=&\frac{i}{Z_H}\sum_{m,n}
e^{-\beta\xi_n}\frac{\langle n|\bJ(\br)|m\rangle\langle m|O|n \rangle}{\xi_m-\xi_n+\omega+i\delta}\\
\nonumber
&& \times \int_0^\beta d\tau e^{(\xi_n-\xi_m)\tau} \\
\nonumber
&=&\frac{i}{\omega+i\delta}\frac{1}{Z_H}\sum_{m,n}\left[e^{-\beta\xi_m}-e^{-\beta(\xi_n-i\epsilon)}\right] \\
\nonumber
&&\times\left[\frac{1}{\xi_m-\xi_n+\omega+i\delta}-\frac{1}{\xi_m-\xi_n+i\epsilon}\right]\\
&&\times\langle n|\bJ(\br)|m\rangle\langle m|O|n \rangle,
\ea
where the limit $\epsilon\rightarrow 0$ is introduced in order to recover the correct result of the $\tau$
integration in the case $\xi_m=\xi_n$, and is to be taken first, followed by the limit
$\delta\rightarrow 0$.

On the other hand, consider the imaginary-time correlation function
\ba
\label{eq:chiLeh}
\nonumber
\Lchi_{O,\bJ}(\br,i\omega_n)&=&-\int_0^\beta d\tau e^{i\omega_n\tau} \langle O(-i\tau)\bJ(\br,0)\rangle \\
\nonumber
&=&\frac{1}{Z_H}\sum_{m,n}\left[e^{-\beta\xi_m}-e^{-\beta(\xi_n-i\epsilon)}\right]\\
&&\times\frac{\langle n|\bJ(\br)|m\rangle\langle m|O|n \rangle}{\xi_m-\xi_n+i\omega_n+i\epsilon},
\ea
where $\omega_n$ is a bosonic Matsubara frequency, and the limit $\epsilon\rightarrow 0$ takes care
of the case $\omega_n=0$ and $\xi_m=\xi_n$. From Eqs. (\ref{eq:PiLeh},\ref{eq:chiLeh}) it then follows that
\be
\label{eq:PiChi}
\Pi_{\bJ,O}(\br,\omega+i\delta)=\frac{i}{\omega+i\delta}\left[\Lchi_{O,\bJ}(\br,\omega+i\delta)
-\Lchi_{O,\bJ}(\br,i\epsilon)\right],
\ee
where $\Lchi$ has been analytically continued via $i\omega_n\rightarrow \omega+i\delta$ to yield the retarded
correlation function.

The above results applies to the calculation of $\hat{\alpha}$ given the identification
$O={\bm J}^e=(1/A)\int d^2r \bJ^e$, $g(\br)=(1/T)\bnabla T\cdot \br$, $Q={\cal K}$, and $\bJ=\bJ^h$. This in turn leads,
together with the definition
\be
\label{eq:defavechi}
\Lchi_{O,{\bm J}}(\omega)=\frac{1}{A}\int d^2r \Lchi_{O,\bJ}(\br,\omega),
\ee
to Eq. (\ref{eq:alpha1}).

From Eq. (\ref{eq:Jedef}) it follows that $\delta\bJ^e=g\bJ^e$, with the consequent
contribution to $\hat\alpha$
\be
\label{eq:alpha2ex1}
\alpha^{(2)}_{ij} = -\frac{1}{AT}\left\langle\int d^2  {\rm J}^e_i(\br) r_j\right\rangle_{\!\! K}.
\ee
To relate it to the $z$ component of the orbital magnetization, $M_z$, note that
\ba
\label{eq:magrel1}
\nonumber
0&=&\left\langle\int d^2r \partial_t\rho^e(\br)r_i r_j\right\rangle_{\!\! K}=
-\left\langle\int d^2r \bnabla\cdot\bJ^e(\br)r_i r_j \right\rangle_{\!\! K} \\
&=&\int d^2r {\rm J}_i^e(\br)r_j + \int d^2r {\rm J}_j^e(\br)r_i,
\ea
where the first equality is a result of $\Tr\{e^{-\beta K}[\rho^e,H]\}=0$, and the third a
result of our assumption $\bJ^e\cdot \bn=0$. Eq. (\ref{eq:magrel1}), together with the definition
\be
\label{eq:Mzdef}
M_z=\frac{1}{2c}\left\langle\int d^2r \left[x{\rm J}^e_y(\br)-y{\rm J}^e_x(\br)\right]\right\rangle_{\!\! K},
\ee
allow us to express $\alpha^{(2)}_{ij}$ by Eq. (\ref{eq:alpha2}).

Next, let us discuss the calculation of $\hat{\tilde{\alpha}}$ using the Kubo formula. Since the calculation
is done for finite $\omega$, which is taken to zero only at the end, one needs to determine the appropriate form
of $\bJ^E$ in the presence of a time-varying electric field. To this end, we split the scalar potential into
$\phi(\br,t)=\phi_0(\br)+\varphi(\br,t)$, such that the driving electric field is given by
$\bE(\br,t)=-\bnabla\varphi(\br,t)-(1/c)\partial_t{\bm A}(\br,t)$, while $\phi_0$(\br) describes
the constant background potential, due to the ions for example. We denote by $\tilde\bJ^E$ the current
density that is given by Eq. (\ref{eq:JEdef}) with time-dependent electromagnetic potentials, and note that it
satisfies $-\bnabla\cdot \tilde\bJ^E(\br,t)=i[H(t),\H(\br,t)]$. Consequently one finds,
\ba
\label{eq:cont-timevar}
\nonumber
\partial_t[\H-\varphi\rho^e]&=&i[H,\H]+\partial_t\H-i\varphi[H,\rho^e]-\partial_t\varphi\rho^e\\
\nonumber
&=&-\bnabla\cdot\left(\tilde\bJ^E-\varphi\bJ^e\right)-\varphi\left(\partial_t\rho^e+\bnabla\cdot\bJ^e\right)\\
\nonumber
&&-\left(\bnabla\varphi+\partial_t{\bm A}/c\right)\cdot\bJ^e\\
&=&-\bnabla\cdot\left(\tilde\bJ^E-\varphi\bJ^e\right)+\bJ^e\cdot\bE,
\ea
which is to be interpreted as a continuity equation, with a source term due to Joule heating, for the
energy density $\rho^E=\H-\varphi\rho^e$, and current $\bJ^E=\tilde\bJ^E-\varphi\bJ^e$ \cite{Aleiner,Karen-kinetic}.
These $\rho^E$ and $\bJ^E$ are also both gauge invariant, with ${\bm J}^E$ obtained from Eq. (\ref{eq:intJE})
via the substitution $\partial_t\rightarrow\partial_t-ie\varphi$.

Introducing the electric field in the gauge $\varphi(\br,t)=-\bE\cdot\br e^{-i(\omega+i\delta)t}$
and applying Eq. (\ref{eq:Kuboeq1}) with $O={\bm J}^h={\bm J}^E+(\mu/e){\bm J}^e$, $g(\br,t)=\varphi(\br,t)$,
$Q=\rho^e$, and $\bJ=\bJ^e$ leads to
\be
\label{eq:alphatilde1}
{\tilde \alpha}_{i j}^{(1)}=\lim_{\omega\rightarrow 0} A\frac{i}{ \omega+ i \delta}
\left[\Lchi_{J^h_i,J^e_j}(\omega+ i\delta)
-\Lchi_{J^h_i,J^e_j}(i\epsilon)\right].
\ee

The above form of the heat current does not change in the presence of $\varphi(\br,t)$. However, in the
limit $\omega\rightarrow 0$, the system relaxes to a state which is close to local (but not global)
thermodynamic equilibrium, for which $\varphi(\br)=-\bE\cdot\br$ becomes a part of the $\phi_0(\br)$
and $\varphi\bJ^e$ a part $\bJ^h$. As a result, an additional contribution to ${\hat{\tilde \alpha}}$ appears,
and is given by
\be
\label{eq:alphatilde2}
{\tilde \alpha}_{i j}^{(2)}=-\frac{1}{A} \left\langle\int d^2r {\rm J}^e_i ({\bf r}) r_j \right\rangle_{\!\! K}
= \frac{c}{A}\epsilon^{ijz}M_z.
\ee
The existence of the magnetization term can be traced back to the assumed local thermodynamic
equilibrium, which implies the relation $T \delta S = \delta E -\mu \delta N +M_z \delta B_z$
for an infinitesimal heat change. This, when divided by $\delta t$ and combined with Faraday's law
$\bnabla\times \bE=-(1/c)\partial_t{\bf B}$, gives Eq. (\ref{eq:alphatilde2}).

Finally, we demonstrate that ${\hat{\tilde\alpha}}^{(1)}$ is gauge invariant. To this end, we employ the
conventional form of the Kubo formula \cite{Mahan}, which for the gauge
$\varphi(\br,t)=-\bE\cdot\br e^{-i(\omega+i\delta)t}$ reads
\ba
\label{eq:Kubo-scalargauge}
\nonumber
\langle {\rm J}_i^h(\br,t)\rangle_{K+\delta K}&=& i\int_{-\infty}^t dt'
\int d^2r' \bE\cdot \br' e^{-i(\omega+i\delta)t'} \\
&&\times\left\langle\left[{\rm J}^h_i(\br,t),\rho^e(\br',t')\right]\right\rangle_{K},
\ea
where we used the fact that in this gauge $\delta{\bm J}^h=0$.

Alternatively, one can use the gauge ${\bf A}(\br,t)={\bf A}^{\rm B}(\br)+{\bf A}^{\rm E}(t)$, where the first
piece is responsible for the magnetic field while the electric field is introduced via
${\bf A}^{\rm E}=-ic\bE e^{-i(\omega+i\delta)t}/(\omega+i\delta)$. The Kubo formula then becomes
\ba
\label{eq:Kubo-timedepgauge}
\nonumber
&&\langle {\rm J}_i^h(\br,t)\rangle_{K+\delta K}=\langle \delta {\rm J}_i^h(\br,t)\rangle_{K}\\
\nonumber
&&+\int_{-\infty}^t dt' \int d^2r' \frac{e^{-i(\omega+i\delta)t'}}{\omega+i\delta}\bE\cdot
\left\langle\left[{\rm J}^h_i(\br,t),\bJ^e_\mu(\br',t')\right]\right\rangle_{K}, \\
\ea
where $K$ includes ${\bf A}^{\rm B}$ but not ${\bf A}^{\rm E}$. To proceed, we note that
\ba
\label{eq:sidecalc}
\nonumber
\int d^2r \bE\cdot \bJ^e &=&\int d\bn\cdot {\tilde{\bf J}}-\int d^2r \,(\bE\cdot \br)\bnabla\cdot \bJ^e \\
&=&\int d^2r\, \bE\cdot\br\, \partial_t\rho^e,
\ea
where in going from the first to the second line we assumed that the surface integral
of $\tilde\bJ=(\bE\cdot\br)\bJ^e$ vanishes. Plugging Eq. (\ref{eq:sidecalc}) into Eq. (\ref{eq:Kubo-timedepgauge})
and integrating by parts over $t'$ yields Eq. (\ref{eq:Kubo-scalargauge}) and a boundary term
\ba
\label{eq:boundaryterm}
\nonumber
&& \frac{e^{-i(\omega+i\delta)t}}{\omega+i\delta}\int d^2r' \bE\cdot\br'
\left\langle\left[{\rm J}^h_i(\br,t),\rho^e(\br',t)\right]\right\rangle_{K} \\
\nonumber
=&&-\frac{e}{c}A^{\rm E}_\nu(t){\bigg \langle}\frac{1}{2 m_i m_\nu}[D_i\psi]^\dagger D_\nu\psi \\
\nonumber
&&+\frac{\delta_{i,\nu}}{2 m_i}\psi^\dagger\left[-\frac{1}{2 m_\nu}D_\nu^2-e\phi
+\frac{1}{2}\int d^2r'U(\br-\br')\rho(\br')\right]\psi \\
\nonumber
&&+\frac{1}{4m_\nu}\int d^2r' d^2r'' \partial_i\left[G(\br,\br')-G(\br,\br'')\right] \\
\nonumber
&&\times \psi^\dagger(\br') \left[\partial'_\nu U(\br'-\br'')\right]
\rho(\br'')\psi(\br')+\delta_{i,\nu}\frac{\mu}{2em_i}\rho^e {\bigg \rangle}_{\!\! K} \\
&&+{\rm H.c.},
\ea
which exactly cancels $\langle \delta {\rm J}_i^h(\br,t)\rangle_{K}$, as can be checked using
Eqs. (\ref{eq:Jedef},\ref{eq:JEdef}). We comment that by applying the considerations outlined
in Appendix \ref{app:timerev} it can be shown that
$\langle \delta {\rm J}_i^h(\br,t)\rangle_{K}$ vanishes in the presence of reflection symmetry.

\section{Behavior of correlation functions under reflection and time reversal}
\label{app:timerev}

We are interested in the case where the potential appearing in the Hamiltonian density (\ref{eq:2dhdensity})
is invariant under reflection about the $y$ axis
\be
\label{eq:parcond1}
\phi(x,y)=\phi(L_x-x,y)\equiv\phi(\tilde x,y),
\ee
and similarly the magnetic field satisfies $B_z(x,y)=B_z(\tilde x,y)$. The latter condition in
obeyed provided that
\ba
\label{eq:parcond2}
A_x(\tilde x,y)=A_x(x,y)+\partial_x f(x,y),\\
A_y(\tilde x,y)=-A_y(x,y)-\partial_y f(x,y).
\ea
It is then straightforward to check that $\pi H(B) \pi^\dagger =H(-B)$ under the reflection transformation
\be
\label{eq:partrans}
\pi\psi(x,y)\pi^\dagger=\psi(\tilde x,y)e^{i(e/c)f(x,y)}.
\ee

As a result, any two bosonic Hermitian operators $O_{1,2}$, transforming according to
\be
\label{eq:paroptrans}
\pi O_{1,2}(B)\pi^\dagger=\epsilon^\pi_{1,2}O_{1,2}(-B),
\ee
with $\epsilon^\pi_{1,2}+\pm 1$, satisfy
\ba
\label{eq:parcorrtrans}
\nonumber
\!\!\!\!\!\!\!\!\!\!\!\!\!\!\left\langle O_1(B) O_2(B)\right\rangle_{\! K(B)}&=&
\Tr\left[\pi e^{-\beta K}O_1 O_2 \pi^\dagger\right]/Z_{K(-B)}\\
&=&\epsilon^\pi_1\epsilon^\pi_2\left\langle O_1(-B) O_2(-B)\right\rangle_{\! K(-B)}.
\ea
Consequently, the imaginary-time correlation function obey
\ba
\label{eq:parfreqcorr}
\nonumber
&&\Lchi_{O_1,O_2}(i\omega_n;B) \\
\nonumber
&&\;\;\;\;\;\;\;\;\;\;\;\;=-\int_0^\beta d\tau e^{i\omega_n\tau}
\left\langle O_1(-i\tau;B) O_2(0;B) \right\rangle_{\! K(B)} \\
&&\;\;\;\;\;\;\;\;\;\;\;\;=\epsilon^\pi_1\epsilon^\pi_2\Lchi_{O_1,O_2}(i\omega_n;-B).
\ea
Note that for both electrical and heat currents $\epsilon_{{\rm J}_{x,y}}^\pi=\mp1$.
Eq. (\ref{eq:parfreqcorr}) also holds for the disorder averaged correlation function
$\overline{\langle O_1 O_2\rangle}_{\! K}=\int D\phi P(\phi) \langle O_1 O_2\rangle_{\! K}$,
even when condition (\ref{eq:parcond1}) is not fulfilled, as long as the disorder distribution
obeys $P[\phi(x,y)]=P[\phi(\tilde x,y)]$.

Under time reversal $\Theta H(B) \Theta^{-1}=H(-B)$. Provided that
\be
\label{eq:timerevoptrans}
\Theta O_{1,2}(B)\Theta^{-1}=\epsilon^\Theta_{1,2}O_{1,2}(-B),
\ee
with $\epsilon^\Theta_{1,2}+\pm 1$, and using
$\langle n|O|n\rangle=\langle \bar n|\Theta O^\dagger \Theta^{-1}|\bar n\rangle$ \cite{Sakurai},
where $|\bar n\rangle = \Theta|n\rangle$ is the time reversed state, one finds
\ba
\label{eq:timerevcorrtrans}
\nonumber
&&\left\langle O_1(-i\tau;B) O_2(0;B)\right\rangle_{\! K(B)} \\
\nonumber
&&\hspace{0.8cm}=\Tr\left[\Theta O_2(0;B) O_1 (i\tau;B)e^{-\beta K(B)}\Theta^{-1}\right]/Z_{K(-B)}. \\
\ea
Hence,
\be
\label{eq:timerevfreq}
\Lchi_{O_1,O_2}(i\omega_n;B)=\epsilon^\Theta_1\epsilon^\Theta_2\Lchi_{O_2,O_1}(i\omega_n;-B).
\ee
For both electrical and heat current densities $\epsilon_{{\rm J}_i}^\Theta=-1$.

\section{Diagonalization of $H_r$}
\label{app:2pdiag}

The relative two-particle Hamiltonian in Eq. (\ref{eq:Hsplit}) is expressed in terms of the operators
\ba
\label{eq:bdef1}
b_1&=&\frac{1}{2^{1/2}}\left(\sqrt{\frac{\Omega}{\omega_c}}\frac{x}{l_x}
+\sqrt{\frac{\omega_c}{\Omega}} l_x\frac{\partial}{\partial x}\right), \\
\label{eq:bdef2}
b_2&=&\frac{1}{2^{1/2}}\left(\sqrt{\frac{\Omega}{\omega_c}}\frac{y}{l_y}
+\sqrt{\frac{\omega_c}{\Omega}} l_y\frac{\partial}{\partial y}\right),
\ea
satisfying $[b_1,b_1^\dagger]=[b_2,b_2^\dagger]=1$, $[b_1,b_2]=[b_1,b_2^\dagger]=0$, as
\ba
\label{eq:Hrdiag1}
\nonumber
H_r&=&\Omega\left(b_1^\dagger b_1+b_2^\dagger b_2 +1\right)
-i\frac{\omega_c}{2}\left(b_1^\dagger b_2-b_2^\dagger b_1\right) \\
&&-\gamma\left[\left(b_1^\dagger+b_1\right)^2-\left(b_2^\dagger+b_2\right)^2\right],
\ea
where
\ba
\label{eq:Omegadef}
\Omega&=&\frac{1}{2}\sqrt{\omega_c^2+\omega_x^2+\omega_y^2} \, , \\
\label{eq:gammadef}
\gamma&=&\frac{\omega_y^2-\omega_x^2}{16\Omega}.
\ea
It may be decoupled into two independent pieces via the canonical transformation
\be
\label{eq:cdef}
\left( \begin{array}{c}
b_1\\
b_2 \\ \end{array}\right) =
\left(\begin{array}{cc}
i\cos\phi & \sin\phi\\
-\sin\phi & -i\cos\phi\\ \end{array}\right) \left(\begin{array}{c}
c_1 \\
c_2\\ \end{array}\right),
\ee
with
\be
\label{eq:phidef}
\tan 2\phi=-\frac{\omega_c}{4\gamma},
\ee
which leads to
\ba
\label{eq:Hdiag2}
\nonumber
H_r&=&\left( \begin{array}{cc}
c_1^\dagger & c_1 \\ \end{array}\right)
\left(\begin{array}{cc}
\Omega_- & \gamma\\
\gamma & \Omega_-\\ \end{array}\right) \left(\begin{array}{c}
c_1 \\
c_1^\dagger\\ \end{array}\right) \\
&+&\left( \begin{array}{cc}
c_2^\dagger & c_2 \\ \end{array}\right)
\left(\begin{array}{cc}
\Omega_+ & -\gamma\\
-\gamma & \Omega_+\\ \end{array}\right) \left(\begin{array}{c}
c_2 \\
c_2^\dagger\\ \end{array}\right),
\ea
where
\be
\label{eq:defOmegapm}
\Omega_\pm=\frac{\Omega}{2}\pm\frac{1}{4}\sqrt{\omega_c^2+(4\gamma)^2}\ge 0.
\ee
Finally, we employ the Bogoliubov transformation
\be
\label{eq:ddef}
\left( \begin{array}{c}
c_{1,2}\\
c_{1,2}^\dagger \\ \end{array}\right) =
\left(\begin{array}{cc}
\cosh\theta_{1,2} & \sinh\theta_{1,2}\\
\sinh\theta_{1,2} & \cosh\theta_{1,2}\\ \end{array}\right) \left(\begin{array}{c}
d_{1,2} \\
d_{1,2}^\dagger\\ \end{array}\right),
\ee
with
\be
\label{eq:theta12def}
\tanh 2\theta_{1,2}=\mp\frac{\gamma}{\Omega_\mp},
\ee
to bring $H_r$ into the diagonalized form
\be
\label{eq:diagHr}
H_r=\omega_1\left(d_1^\dagger d_1+\frac{1}{2}\right)+\omega_2\left(d_2^\dagger d_2+\frac{1}{2}\right),
\ee
where $[d_1,d_1^\dagger]=[d_2,d_2^\dagger]=1$, $[d_1,d_2]=[d_1,d_2^\dagger]=0$, and the eigenfrequencies
are given by
\be
\label{eq:omega12def}
\omega_{1,2}=2\sqrt{\Omega_\mp^2-\gamma^2}.
\ee

The center of mass part of the two particle eigenstate is obviously symmetric under particle exchange.
To maintain antisymmetry of the full state we require that the relative part would be antisymmetric.
One can check that the wavefunction of the ground state, $|n_1=0,n_2=0\rangle$, of $H_r$ is proportional
to $\exp[-(ax^2+ibxy+cy^2)]$, with $a,b,c$ constants, and hence symmetric.
From Eqs. (\ref{eq:bdef1},\ref{eq:bdef2}) it follows that $b_{1,2}-\rightarrow-b_{1,2}$ under particle
exchange, and the linearity of the ensuing transformations means that $d_{1,2}$ share this property.
Thus, the allowed $|n_1,n_2\rangle\propto (d_1^\dagger)^{n_1}(d_2^\dagger)^{n_2}|0,0\rangle$ are
those for which $n_1+n_2$ is odd.

\section{Calculating $\hat{\alpha}$ for the quasi-one-dimensional model}
\label{app:1dalpha}

\subsection{Calculating $\alpha_{yx}^{(1)}$}
\label{app:1dalphayx1}

According to Eq. (\ref{eq:alpha1}), $\alpha_{yx}^{(1)}$ is determined from the correlation function
$\Lchi_{J_y^e,J_x^h}(i\omega_n)$, which we evaluate perturbatively in $H_{\sJ}$. One can readily verify
that the zeroth order term vanishes in the limit $L\rightarrow\infty$, and the lowest non-vanishing contribution is
\be
\label{eq:chiyx1d1}
\Lchi_{J_y^e,J_x^h}(i\omega_n)=\int_0^\beta d\tau d\tau' e^{i\omega_n\tau}\langle T_\tau J_y^e(\tau)J_x^h(0)H_{\sJ}(\tau')
\rangle_0,
\ee
where here $O(\tau)=e^{H_0\tau}Oe^{-H_0\tau}$ and $T_\tau$ is the imaginary time ordering operator. Using expressions
(\ref{eq:Jey1d}) and (\ref{eq:Jhx1d}) of the current densities and the averages \cite{Giamarchi-book}
\ba
\label{eq:F1}
\nonumber
F_1(x,\tau)&=&K^{-1}\langle T_\tau\left[\phi(x,\tau)-\phi(0,0)\right]^2\rangle_0 \\
\nonumber
&=&K\langle T_\tau\left[\theta(x,\tau)-\theta(0,0)\right]^2\rangle_0 \\
\nonumber
&=&\frac{1}{2\pi}\ln\left\{\left(\frac{l_T}{a}\right)^2\!\left[\sinh^2\left(\frac{x}{l_T}\right)
+\sin^2\left(\frac{v\tilde\tau}{l_T}\right)\right]\right\},\\
\\
\label{eq:F2}
\nonumber
F_2(x,\tau)&=&\langle T_\tau\phi(x,\tau)\theta(0,0)\rangle_0 \\
\nonumber
&=&\frac{1}{4\pi}\left\{\ln\left[-i\sinh\left(\frac{x+iv\tilde\tau}{l_T}\right)\right]\right. \\
&&\left.-\ln\left[i\sinh\left(\frac{x-iv\tilde\tau}{l_T}\right)\right]\right\},
\ea
where $\tilde\tau=\tau+{\rm sign}(\tau)a/v$, and $l_T=v/\pi T$, we obtain
\ba
\label{eq:chiyx1d2}
\nonumber
\Lchi_{J_y^e,J_x^h}(i\omega_n)&=&\frac{2\pi ev^2\J^2}{KA}\int_0^\beta d\tau d\tau'\int dx dx' e^{i\omega_n\tau}\\
\nonumber
&&\times C(x-x',\tau-\tau')\sin[b(x-x')]\\
\nonumber
&&\times \left[\partial_xF_1(x,\tau)-\partial_{x'}F_1(x',\tau')\right]\\
&&\times \left[\partial_xF_2(x,\tau)-\partial_{x'}F_2(x',\tau')\right],
\ea
with
\be
\label{eq:Cdef}
C(x,\tau)=e^{-2\pi K^{-1}F_1(x,\tau)}.
\ee
The parity of the functions $F_1$ and $F_2$ leads, after defining $r=x-x'$, to
\ba
\label{eq:chiyx1d3}
\nonumber
\Lchi_{J_y^e,J_x^h}(i\omega_n)&=&\frac{2\pi ev^2\J^2}{KA}\int_0^\beta d\tau d\tau'\int dx dr e^{i\omega_n\tau}\\
\nonumber
&&\times C(r,\tau-\tau')\sin(br)\\
\nonumber
&&\times \left[\partial_xF_1(x,\tau)\partial_{x}F_2(x-r,\tau')\right.\\
&&+\left.\partial_xF_1(x-r,\tau')\partial_{x}F_2(x,\tau)\right].
\ea
For $\tau,\tau'\in[0,\beta]$ the integral over $x$ can be evaluated with the result
\ba
\label{eq:chiyx1d4}
\nonumber
\!\!\!\!\!\!\!\!\!\!\!\!\!\Lchi_{J_y^e,J_x^h}(i\omega_n)&=&\frac{i ev\J^2}{\pi l_T A}\int_0^\beta d\tau d\tau'\int dr
e^{i\omega_n\tau}\sin(br)\\
\!\!\!\!\!\!\!\!\!\!\!\!\!&&\times \left[v^2(\tau-\tau')\partial_r + r\partial_{\tau'}\right]C(r,\tau-\tau').
\ea
Integrating by parts, we find that the $\partial_{\tau'}$ term vanishes. Finally, integration by parts over $r$ and
a change of variables to $\tau\pm\tau'$, gives
\be
\label{rq:chiyx1dfinal}
\Lchi_{J_y^e,J_x^h}(i\omega_n)=\frac{ev^2\J^2 b}{\omega_n A}\left[C(b,i\omega_n)-C(b,0)\right],
\ee
where $\Lchi_{J_y^e,J_x^h}(i\omega_n=0)=0$, and
\be
\label{eq:Cqomegadef}
C(q,i\omega_n)=\int_{-\infty}^\infty dx\int_0^\beta d\tau\, e^{i(\omega_n\tau-qx)}C(x,\tau).
\ee

\vspace{-0.5cm}
\begin{figure}[t!!!]
\centering
  \includegraphics[width=\linewidth,clip=true]{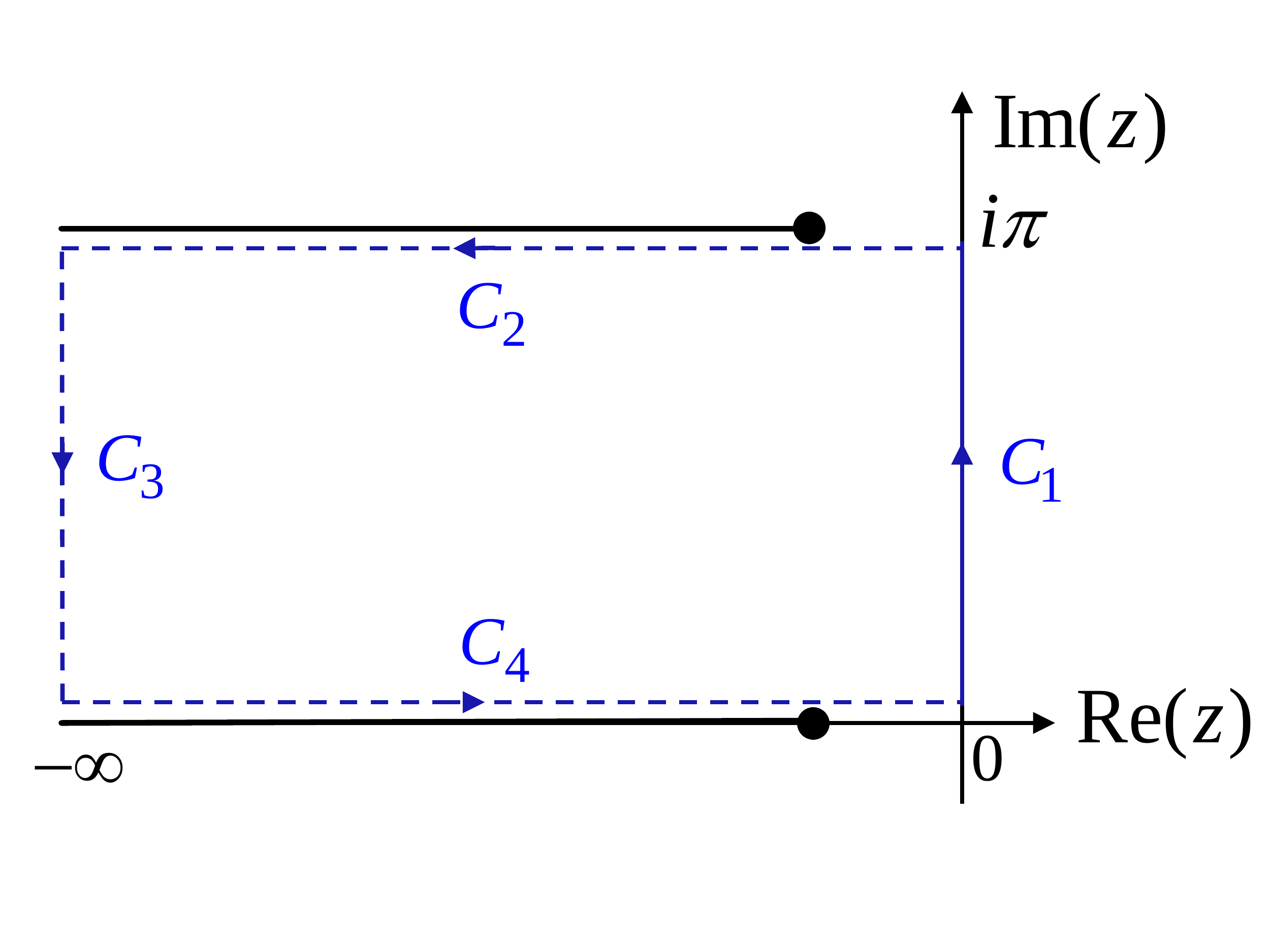}
  \caption{The integration contour for calculating $C(q,i\omega_n)$.}
  \label{fig:Ccontour}
\end{figure}

We are now left with the task of calculating
\ba
\label{eq:Cqomegaint1}
\nonumber
C(q,i\omega_n)&=&\frac{l_T^2}{v}\left(\frac{a}{l_T}\right)^{2/K}\int_{-\infty}^\infty dx \int_0^\pi dy \,
e^{i(\tilde\omega_n y- \tilde q x)}\\
&&\times (\sinh^2 x+\sin^2 y)^{-1/K},
\ea
with $\tilde\omega_n=l_T\omega_n/v$ and $\tilde q=q l_T$. Clearly, this function is symmetric under
$\omega_n\rightarrow-\omega_n$, so in the following we assume $\omega_n>0$. Next, a change in the integration variable
to $z=iy$ rotates the integral onto the segment $C_1$, see Fig. {\ref{fig:Ccontour}}. Applying
Cauchy's theorem, while noting the branch cuts at ${\rm Re}(z)<0$ and ${\rm Im}(z)=0,\pi$, we may trade $C_1$ by the
contour $-C_2-C_4$. Finally, we use the invariance of the integrand under $z\rightarrow z-i\pi$ to shift $C_2$ below the
real axis and obtain
\ba
\label{eq:Cqomegaint2}
\nonumber
C(q,i\omega_n)&=&i\frac{l_T^2}{v}\left(\frac{a}{l_T}\right)^{2/K}\int_{-\infty}^\infty dx \int_{-\infty}^0 dz \,
e^{\tilde\omega_n z- i\tilde q x}\\
\nonumber
&&\times \left\{\left[\sinh^2 x-\sinh^2(z-i\epsilon)\right]^{-1/K}\right. \\
&&\left.-\left[\sinh^2 x-\sinh^2(z+i\epsilon)\right]^{-1/K}\right\},
\ea
where $\epsilon$ is a positive infinitesimal. Taking $z\rightarrow -z$ and using
\ba
\label{eq:sinhiden}
\nonumber
&&\left[\sinh^2 x -\sinh^2(z\pm i\epsilon)\right]^{1/K}\\
\nonumber
&&=\left\{\begin{array}{cc}
(\sinh^2x-\sinh^2z)^{1/K} & |x|>|z|\\
(\sinh^2z-\sinh^2x)^{1/K}e^{\mp i\pi{\rm sign(z)}/K} & |z|>|x| \\ \end{array}\right., \\
\ea
we arrive at
\ba
\label{eq:Cqomegaint3}
\nonumber
\!\!\!\!\!\!\!\!\!\!\!\!\!\!\!
C(q,i\omega_n)&=&2\frac{l_T^2}{v}\left(\frac{a}{l_T}\right)^{2/K}\sin\left(\frac{\pi}{K}\right) \\
\!\!\!\!\!\!\!\!\!\!\!\!\!\!\!
&\times&\int_0^\infty dz \int_{-z}^z dx\frac{e^{-\tilde\omega_n z- i\tilde q x}}{\left(\sinh^2z-\sinh^2x\right)^{1/K}}.
\ea
By changing variables to $z\pm x$ the remaining integrals can be evaluated for $K>1$. The result,
after analytically continuing $i\omega_n\rightarrow \omega+i\delta$, is given by Eq. (\ref{eq:Cqomega}).

\subsection{Calculating $\alpha_{xy}^{(1)}$}
\label{app:1dalphaxy1}

Within a perturbative treatment of $\J$ the leading contribution to $\alpha_{xy}^{(1)}$ is determined by
\be
\label{eq:chixy1d1}
\Lchi_{J_x^e,J_y^h}(i\omega_n)=\int_0^\beta d\tau d\tau' e^{i\omega_n\tau}\langle T_\tau J_x^e(\tau)J_y^h(0)H_{\sJ}(\tau')
\rangle_0.
\ee
Concentrating on the spatial integrals which appear in this contribution, we find that it is
proportional to
\ba
\label{eq:chixy1dint}
\nonumber
&&\frac{1}{A N_c L}\sum_{j=1}^{N_c}\sum_{j'=2}^{N_c}\int_{-L/2}^{L/2} dx dx' dx''
\sin[b(x'-x'')]\\
\nonumber
&&\times C(x'-x'',\tau'-\tau'')\left\{\delta_{j,j'}\partial^2_x F_2(x-x',\tau'-\tau)\right. \\
\nonumber
&&+(\pi/K)(\delta_{j,j'}-\delta_{j,j'-1}) \partial_{x'} F_2(x'-x'',\tau'-\tau'') \\
&&\times[\partial_{x'}F_1(x'-x,\tau'-\tau)-\partial_{x''}F_1(x''-x,\tau''-\tau)]\}.\;\;\;\;\;\;\;\;\;
\ea
Clearly, the sum over $j'$ of the last two lines vanishes. The sums and $x$ integral over the remaining
part give $N_c \partial_x F_2(x-x',\tau-\tau')|_{x=-L/2}^{x=L/2}$. It follows from Eq. (\ref{eq:F2}) that
this term is appreciable only for $x'$ within a distance of order $l_T$ from the the edges at $\pm L/2$.
Consequently, we conclude that $\alpha_{xy}^{(1)}$ is smaller by a factor $l_T/L$ than the
corresponding $\alpha_{yx}^{(1)}$.

\subsection{Calculating $M_z$}
\label{app:1dMz}

The magnetization can be computed from the thermodynamic relation
\be
\label{eq:magterno}
M_z=-\left(\frac{\partial\Omega}{\partial B}\right)_{\mu,T},
\ee
where $\Omega$ is the grand canonical potential. To second order in $\J$ we
obtain
\ba
\label{eq:Omegaexp}
\nonumber
\Omega&=&\Omega_0-\frac{1}{2}\int_0^\beta d\tau \langle T_\tau H_{\sJ}(\tau)H_{\sJ}(0)\rangle_0 \\
&=&\Omega_0-\frac{\J^2A}{4d}\int_0^\beta d\tau \int dx  C(x,\tau)\cos(bx), \;\;\;\;\;
\ea
with $\Omega_0=-T{\rm Tr}[e^{-\beta H_0}]$, from which it follows, using Eqs. (\ref{eq:magterno}) and
(\ref{eq:Cqomegadef}), that
\be
\label{eq:Magres}
M_z=-\frac{e\J^2A}{2c}\frac{\partial C(b,0)}{\partial b}.
\ee

The same result is also obtained from the definition of $M_z$ in terms of currents, Eq. (\ref{eq:Mzdef}).
To see this we note that Eqs. (\ref{eq:1dHJ}) and (\ref{eq:Jey1d}) imply
\ba
\label{eq:magcurrenty}
\nonumber
\left\langle\int d^2 r x{\rm J}^e_y\right\rangle_{\!\!H}&=&-2ed\left\langle\frac{\partial H_{\sJ}}{\partial b}
\right\rangle_{\!\!H}\\
\nonumber
&=&ed \frac{\partial}{\partial b}\int_0^\beta d\tau \left\langle T_\tau H_{\sJ}(\tau)H_{\sJ}(0)\right\rangle_0 \\
&=&cM_z.
\ea
Furthermore, explicit calculation reveals that to order $\J^2$
\ba
\label{eq:magcurrentx}
\nonumber
\!\left\langle\int d^2 r y{\rm J}^e_x\right\rangle_{\!\!H}&=&
-\frac{evd\J^2}{2} \sum_{j=1}^{N_c}\sum_{j'=2}^{N_c}j\left(\delta_{j,j'}-\delta_{j,j'-1}\right) \\
\nonumber
&\times&\int_0^\beta d\tau d\tau' \int dx dx' dx'' \sin[b(x'-x'')] \\
\nonumber
&\times& C(x'-x'',\tau'-\tau'')\partial_x F_1(x-x',\tau-\tau'). \\
\ea
A naive evaluation of the integral, disregarding the finite size of the system, would yield zero owing to
the fact that the integrand is odd in $x-x'$ and $x'-x''$. However, a more careful analysis leads to a
different conclusion. First, the sums add up to $N_c-1$. Secondly, Eq. (\ref{eq:F1}) implies that to a
good approximation $F_1(\pm L/2-x',\tau-\tau')=[\ln(l_T/2a)+(L/2\mp x')/l_T]/\pi$, as long as $x'$ is
situated more than $l_T$ away from the edges. Using this we obtain
\ba
\label{eq:magcurrentx2}
\nonumber
\left\langle\int d^2 r y{\rm J}^e_x\right\rangle_{\!\!H}&=&
\frac{e\J^2A}{2} \int_0^\beta d\tau \int dx C(x,\tau)\, x \sin(bx) \;\; \\
&=&c\frac{\partial\Omega}{\partial B}=-cM_z,
\ea
as required.

\end{document}